\documentclass[%
 reprint,
 amsmath,amssymb,
 aps,
]{revtex4-1}
\usepackage{graphicx}
\usepackage{dcolumn}
\usepackage{bm}

\begin{document}


\title{Slow quenches in two-dimensional time-reversal symmetric  $\mathbb{Z}_2$ topological insulators}

\author{Lara Ul\v{c}akar}

\affiliation{Jozef Stefan Institute, Jamova 39, Ljubljana, Slovenia}

\email{lara.ulcakar@ijs.si}

\author{Jernej Mravlje}

\affiliation{Jozef Stefan Institute, Jamova 39, Ljubljana, Slovenia}

\affiliation{Faculty for Mathematics and Physics, University of Ljubljana, Jadranska
	19, Ljubljana, Slovenia}

\author{Anton Ram\v{s}ak}

\affiliation{Jozef Stefan Institute, Jamova 39, Ljubljana, Slovenia}

\affiliation{Faculty for Mathematics and Physics, University of Ljubljana, Jadranska
	19, Ljubljana, Slovenia}

\author{Toma\v{z} Rejec}

\affiliation{Jozef Stefan Institute, Jamova 39, Ljubljana, Slovenia}

\affiliation{Faculty for Mathematics and Physics, University of Ljubljana, Jadranska
	19, Ljubljana, Slovenia}

\date{\today}	

\begin{abstract}
We study the topological properties and transport in the Bernevig-Hughes-Zhang (BHZ) model undergoing a slow quench between different topological regimes. Due to the closing of the band gap during the quench, the system ends up in an excited state. For quenches governed by a Hamiltonian that preserves the symmetries present in the BHZ model (time-reversal, inversion, and conservation of spin projection), the $\mathbb{Z}_2$ invariant remains equal to the one evaluated in the initial state. The bulk spin Hall conductivity does change and its time average approaches that of the ground state of the final Hamiltonian. The deviations from the ground-state spin Hall conductivity as a function of the quench time follow the Kibble-Zurek scaling. We also consider the breaking of the time-reversal symmetry, which restores the correspondence between the bulk invariant and the transport properties after the quench.

\end{abstract}

\pacs{71.10.Pm, 03.65.Vf, 73.43.-f, 68.65.Fg}
\maketitle


\section{\label{sec:Intro}Introduction}

Topological insulators have been one of the focal points of condensed matter physics for the last decade due to their interesting properties and potential future applications in nano-electronics \cite{MemCell,MagSwitch,Thermoel,ElInterconn,AppRev}. The topological insulators are bulk insulators that at the edges host gapless conducting states which avoid dissipation \cite{Klitzing80,Laughlin81,Thouless82,KaneMele}.
While the ground-state physics of topological insulators is already well established, less is known about their response to the time-dependent driving, which is a subject of an active current investigation. 

Recently, the question of the response due to the changes of system's parameters (quantum quench)  was explored in Chern insulators. Their topological phase is characterized by a non-trivial Chern number and the quantum Hall effect. By definition, the topological invariants are conserved by any adiabatic evolution. But, strikingly, a stronger statement holds. The Chern number is conserved for an arbitrary evolution (the only restriction that the Hamiltonian is smooth in momentum) even if during the evolution the band gap closes~\cite{PxPySuperFluid,Rigol2015}. Several  works \cite{Caio2015,Caio2016,Wilson16,WangNJP16,Wang16,Duta17} investigated response due to a rapid change of parameters (i.e., a sudden quench) and found that the bulk Hall response, in contrast to the Chern number, evolves in time.  Refs. \onlinecite{zoller,Unal16,Schuler17} studied a different case in which the parameters are varied slowly between different topological regimes (i.e., a slow quench). The subtle point is that even though the change of parameters is slow, it can never be adiabatic as the band gap closes. For slow quenches, the bulk Hall response of the post-quench system was found to approach the value determined by the topological invariant of the ground state of the final Hamiltonian. In a more general setting, response due to slow quenches was also studied in topological superconductors \cite{Bermudez10,DeGottardi11,PxPySuperFluid,Sacramento14}.

Topological systems are gapless at the critical point between different topological regimes, thus even slow quenches between different topological regimes create excitations. The Kibble-Zurek argument \cite{Kibble,Zurek} predicts that the density of excitations scales with the quench time to the power given by the critical exponents, associated with the critical point across which the system is quenched. Agreement of the Kibble-Zurek scaling and the actual response of the bulk after the quench was shown in Refs.~\onlinecite{Dutta10,Privitera16} for Chern insulators and in Ref.~\onlinecite{Bermudez10} for one-dimensional superconductors.

It is natural to ask whether the above findings are general and apply to other types of topological insulators. We consider two-dimensional topological insulators with the time-reversal symmetry (TRS). They are characterized by the $\mathbb{Z}_2$ topological invariant and, in the topological phase, exhibit the spin Hall effect. Importantly, several experimental realizations of these systems exist \cite{BHZmodel,HgTeExp,InAsQW,InAsExp,Bismuthene,HighTempTop}.

In this paper we study the topological invariant and the spin Hall effect of the BHZ model. We set the system to the ground state of the Hamiltonian in a certain topological regime and then slowly quench the Hamiltonian to a different topological regime. Despite the fact that the Hamiltonian is time-reversal symmetric at all times, the TRS of the post-quench state is broken \cite{McGinleyCooper}. Correspondingly, in the general case with only TRS, the post-quench $\mathbb{Z}_2$ invariant is ill-defined. However, for systems described by the BHZ model, which are inversion symmetric and conserve the  spin projection $s_z$, the $\mathbb{Z}_2$ invariant (Eq.~\eqref{Z2def}) remains well defined and does not change during the quench.

In the limit of an infinitely slow quench the bulk spin Hall conductivity approaches the value characteristic of the ground state of the final Hamiltonian. We show that the deviations from the ground-state value obey the Kibble-Zurek scaling. We also explore quenches that break the TRS of the Hamiltonian and keep the band gap open at all times (such quenches cannot be realized for Chern insulators). In this case the correspondence is restored: the topological invariant of the system and the spin Hall response are both characteristic of the ones evaluated for the ground state of the final Hamiltonian.

The paper is structured as follows. In Sec.~\ref{sec:model} we introduce the BHZ model and describe how we evaluate  the $\mathbb{Z}_2$ invariant and the spin Hall conductivity. Sec.~\ref{sec:results} is dedicated to the results: the phase diagram of the BHZ model is explored and the response of the system after a quench that preserves the TRS of the Hamiltonian as well as after a quench that breaks the TRS of the Hamiltonian are shown. In Sec.~\ref{sec:concl} our findings are summarized. In Appendix \ref{app:bandProp}, band dispersions and post-quench occupancies are analytically discussed. We show the spin Berry curvature \cite{SpinBerryCurv} and analyse the deviation of the post-quench spin Hall conductivity from the ground state value in detail. We also calculate the critical exponents of the BHZ model. In Appendix~\ref{app:proofs} we generalize the proof of the conservation of the Chern number to the case of multiple band systems and we discuss the behaviour of the $\mathbb{Z}_2$ invariant after the quench. We present the full formula for the spin Hall conductivity and we discuss the oscillations of the spin Hall response of the post-quench system in Appendix~\ref{app:perturb}.

\section{\label{sec:model}Model and methods}

We study a two dimensional time-reversal symmetric $s=1/2$ system on a square crystalline lattice with two orbitals per unit cell. Its bulk momentum-space Hamiltonian is given by
\begin{equation}
\begin{split}
\hat{H}(\mathbf{k})=\hat{s}_0 \otimes [(u+\cos{k_x}+\cos{k_y})\hat{\sigma}_z+\sin{k_y}\hat{\sigma}_y]\\
+\hat{s}_z \otimes \sin{k_x} \hat{\sigma}_x +c\,\hat{s}_x \otimes \hat{\sigma}_y,\label{eq:H}
\end{split}
\end{equation}
where $\hat{s}_i$ and $\hat{\sigma}_i$ for $i\in \{x,y,z\}$ are Pauli operators, $\hat{s}_0$ and $\hat{\sigma}_0$ are identity operators in spin space and local orbital space, respectively, and $\mathbf{k}$ is an element of the first Brillouin zone (BZ). $c\in \mathbb{R}$ is the coupling constant between spin and orbital degrees of freedom and $u$ is the staggered orbital binding energy. The Hamiltonian is expressed in the units of inter-cell hopping amplitude which is equal in both $x$ and $y$ directions. We also set $\hbar$ and the lattice constant to $1$. The system has the TRS with the time-reversal operator $\hat{\mathcal{T}}=i\hat{s}_y K$, $K$ being the complex conjugation. When  $c=0$, the original BHZ model \cite{BHZmodel} is recovered, in which the perpendicular projection of the spin $s_z$ is conserved. It describes the low-energy physics of the HgTe/CdTe quantum wells. In systems with  band inversion asymmetry and structural inversion asymmetry, such as InAs/GaSb/AlSb Type-II semiconductor quantum wells \cite{InAsQW}, terms that couple states with opposite spin projections and preserve the TRS arise. We model such terms with the simplified $c\neq 0$ term.
We consider half-filled systems at zero temperature, meaning that before the quench the lower two energy bands are occupied and the upper two are empty.

We describe the time-reversal symmetric insulator by a set of occupied states $\{|u_n(\mathbf{k})\rangle,\,\mathbf{k}\in \mathrm{BZ},\,1\le n\le N_F\}$. A phase of the bulk time-reversal symmetric insulator is characterized by the $\mathbb{Z}_2$ invariant $N_\mathrm{bulk}$ that distinguishes between the topological regime ($N_\mathrm{bulk}=1$) and the trivial band insulator regime ($N_\mathrm{bulk}=0$). In numerical evaluation it is convenient to use a gauge invariant definition of the $\mathbb{Z}_2$ invariant $N_\mathrm{bulk}$: it is equal to the parity of the number of times the Wannier centre flow $\theta_n (k_y)$, in range $k_y \in (0,\pi)$, crosses an arbitrarily chosen fixed value $\tilde{\theta}\in [-\pi,\pi)$ \cite{WannCentre},
\begin{equation}
N_n (\tilde{\theta})=\mathrm{number}\,\mathrm{of}\,\mathrm{solutions}\,k_y \in (0,\pi) \,\mathrm{of}\,\theta _n (k_y)=\tilde{\theta},
\end{equation}
\begin{equation}
N_{\mathrm{bulk}}=\left(\sum_{n=1}^{N_F}N_n (\tilde{\theta})\right) \mathrm{mod}\,2.\label{Z2def}
\end{equation}
The Wannier centre flow $\theta_n (k_y)$ is equal to the phase of the $n$-th eigenvalue of the Wilson loop, a multi-band generalization of the Berry phase. The Wilson loop is defined as
\begin{equation}
W(k_y)=M^{(12)}M^{(23)}\ldots M^{(N-1,N)}M^{(N,1)},
\end{equation}
\begin{equation}
M^{(kl)}_{nm}=\langle u_n (k \delta_k -\pi,k_y) | u_m (l \delta_k -\pi,k_y) \rangle,
\end{equation}
where $\delta_k=2\pi/N$ is the discretization step in the momentum space of a lattice with periodic boundary conditions and $N\times N$ sites. Matrices $W(k_y)$ and $M^{(kl)}$ are of dimension $N_F\times N_F$. Wannier centre flow can be associated with the expectation value of the relative position of a state from the nearest lattice site. The $\mathbb{Z}_2$ invariant is well defined only for systems in which the Wannier centre flow is symmetric about and doubly degenerate at $k_y=0,\pi$. For pedagogical discussion see Ref.~\onlinecite{madzari}.

We evaluate the  spin Hall conductivity $\sigma_{xy}^{\mathrm{spin}}$ by calculating the spin current density $j_y ^\mathrm{spin}$ as a response to an electric field $E_x$ in the perpendicular direction. 
For the spin current defined as \cite{SHallRashba,SpinBerryCurv} (see Refs.~\onlinecite{Murakami,SpinCurrentSOI,OnsagerRel} for other possible definitions)
\begin{equation}
\hat{j}_y^{\mathrm{spin}}= \frac{1}{2} \frac{1}{N^2} \, \hat{s}_z\, \frac{\partial \hat{H}(\mathbf{k})}{\partial k_y},\label{eq:spinCurrent}
\end{equation}
the spin Hall conductivity can be evaluated as
\begin{equation}
\sigma_{xy}^\mathrm{spin}=\frac{1}{E_x}\sum_{n=1} ^{N_F}\, \sum_{\mathbf{k}}\langle u_n (\mathbf{k})| \hat{j}_y^{\mathrm{spin}}|u_n (\mathbf{k})\rangle.\label{eq:SpinHallConduct}
\end{equation}
$E_x$ is a small homogeneous electric field  switched on at $t=t_E$,  $E_x(t)=E_0[1-\exp (-(t-t_E)/\tau_E)]$. Throughout the paper we choose $\tau_E=10$, $E_0=0.0001$ and the system size $200\times 200$. We checked that increasing the system size further does not affect the results presented in the paper. To preserve the translational symmetry we introduce the electric field through a spatially homogeneous time-dependent vector potential $A_x(t)=-\intop E_x(t)\mathrm{d}t$ \cite{OnsagerRel}.

\section{\label{sec:results}Results}
\subsection{\label{sec:PhaseDiag}Phase diagram of the BHZ model}

\begin{figure*}
	\centerline{\includegraphics[width=145pt]{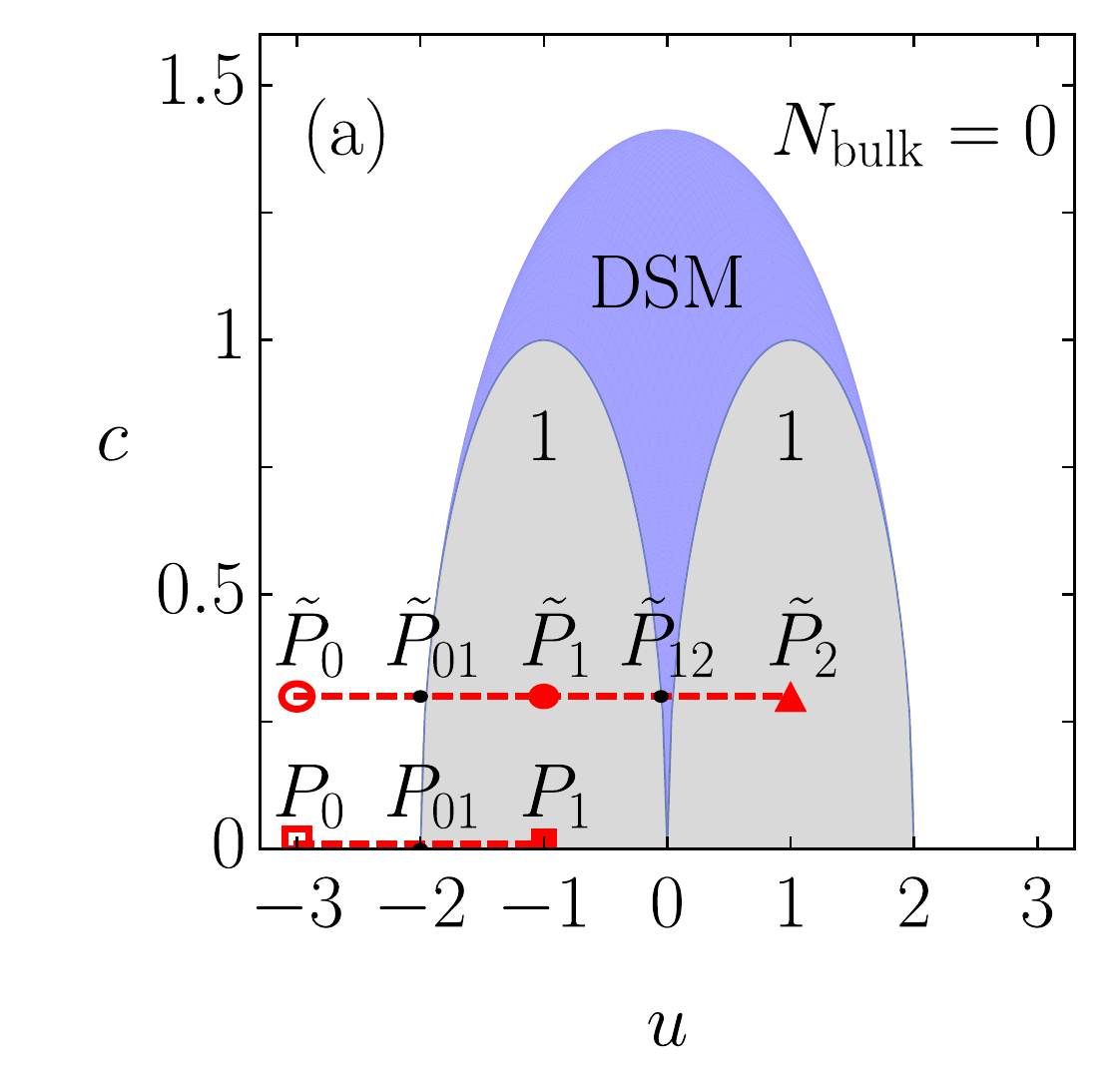} \includegraphics[width=145pt]{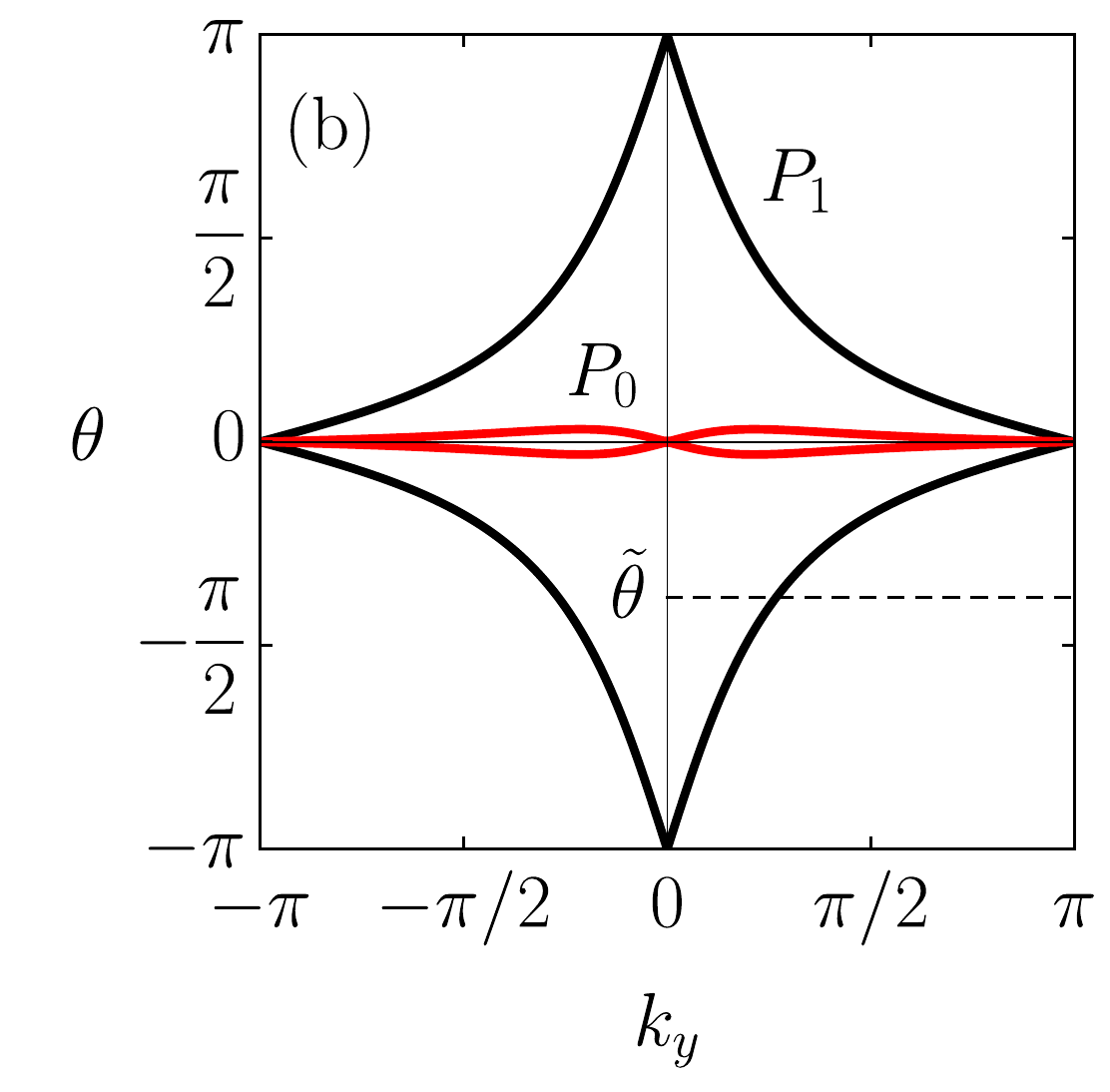}\includegraphics[width=145pt]{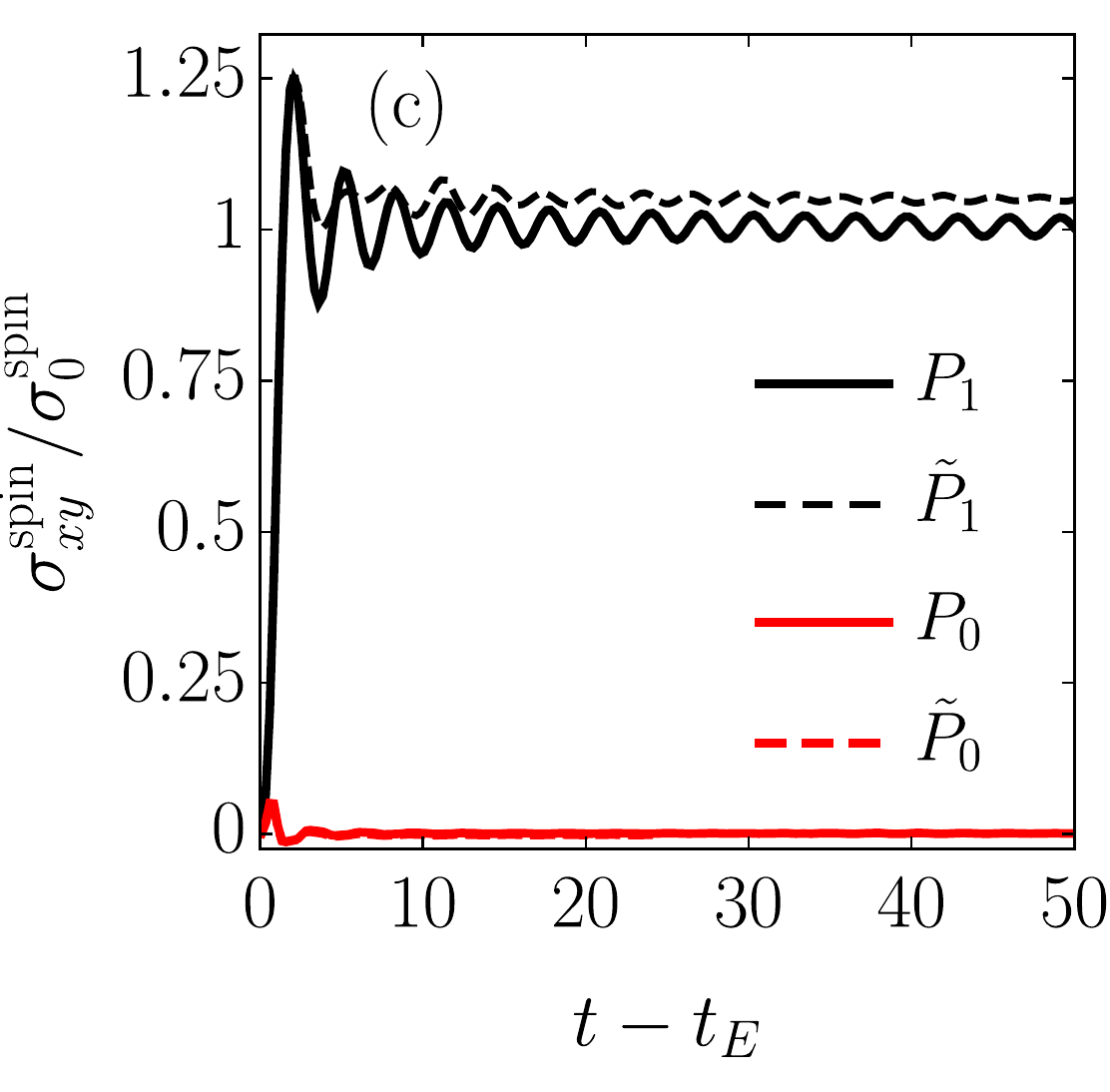}}
	\caption{ (a) Phase diagram of the BHZ model in $P=(u,c)$ parameter space. The grey coloured area marks topological insulator parameter regimes where $N_\mathrm{bulk}=1$, white area the trivial insulator regime with $N_{\mathrm{bulk}}=0$ and  the blue area the Dirac semimetal regime. Points $P_0=(-3,0)$, $\tilde{P}_0=(-3,0.3)$, $P_1=(-1,0)$, $\tilde{P}_1=(-1,0.3)$ and $\tilde{P}_2=(1,0.3)$ mark initial and final points between which parameters of the Hamiltonian are quenched (red, dashed). Several band gap closing points are marked as $P_{01}$, $\tilde{P}_{01}$ and $\tilde{P}_{12}$. (b) Wannier centre flows for the ground state of the Hamiltonian at $P_0$ (red) and  $P_1$ (black). The dashed black line presents a chosen $\tilde{\theta}$ used for the determination of the $\mathbb{Z}_2$ invariant. (c) Spin Hall conductivities at $P_0$ (red), $\tilde{P}_0$ (red, dashed), $P_1$ (black), $\tilde{P}_1$ (black, dashed), where $\sigma_0^\mathrm{spin}=e/2\pi$.}
	\label{fig:PhaseDiagram}
\end{figure*}

The phase diagram of the  BHZ model, shown in Fig.~\ref{fig:PhaseDiagram} (a), describes the topological phase of the ground state of the Hamiltonian \eqref{eq:H} at parameters $P=(u,c)$. It consists of three insulating regions: the trivial regime with $N_{\mathrm{bulk}}=0$ (white) and two topological regimes with $N_{\mathrm{bulk}}=1$ (grey). Insulating regimes are separated from each other by a broad Dirac semimetal regime (blue region) in which the system has a closed band gap with linear dispersion. The upper boundary between the semimetal and the trivial insulator regimes is $c(u)=\sqrt{2-u^2/2}$, while the lower boundaries between the topological insulator and semimetal regimes are $c(u)=\sqrt{1-(u-1)^2}$ and $c(u)=\sqrt{1-(u+1)^2}$. In the semimetal regime the band gap closes at different points in the Brillouin zone, in particular for $P_{01}=(-2,0)$ at the $\Gamma$ point at $\mathbf{k}=(0,0)$, for $P=(0,0)$ at $\mathbf{k}=(0,\pi)$ and $(\pi,0)$, for $P=(2,0)$ at $\mathbf{k}=(\pi,\pi)$ and for $P=(\pm 1, 1)$ at $\mathbf{k}=(\pm \pi/2,0)$ and $(0,\pm \pi/2)$. In Appendix~\ref{app:bandProp}, analytical expressions for the band dispersions near the band gap closing are given and graphs are shown at the band gap closing points $P_{01}$, $\tilde{P}_{01}$ and $\tilde{P}_{12}$.

For the purpose of later comparison with the response after the quench, we first establish what kind of behaviour to expect in distinct parameter regions for the ground state. Fig.~\ref{fig:PhaseDiagram} (b) shows Wannier centre flows $\theta(k_y)$ at $P_0$ (red) and $P_1$ (black) while the dashed line is an arbitrary $\tilde{\theta}$ chosen to evaluate the Eq.~\eqref{Z2def}. For instance, one can see that in the ground state at $P_1$ the Wannier centre flow crosses the dashed line once, hence $P_1$ corresponds to topological regime. 

The spin Hall conductivity evaluated as described in Eq. \eqref{eq:SpinHallConduct} is presented in Fig.~\ref{fig:PhaseDiagram} (c). One can see a sharp distinction between the result in the topological regime ($P_1$ and $\tilde{P}_1$) where, following a steep rise, the spin Hall conductivity oscillates around a finite value and the trivial regime ($P_0$ and $\tilde{P}_0$) where it oscillates around zero instead. The frequency of these oscillations is equal to the band gap. The amplitude of the oscillations diminishes with time and it also becomes smaller if the electric field is turned on more adiabatically (i.e., with longer $\tau_E$).  In the topological regime, the ground-state value of the spin Hall conductivity is quantized in the units of $\sigma_0^\mathrm{spin}=e/2\pi$ ($e$ is the charge of electric carriers) whenever $s_z$ is conserved (e.g, in $P_1$ it is equal to $\sigma_0^\mathrm{spin}$), but has a non-quantized value elsewhere \cite{KaneMele}. In contrast to the spin Hall conductivity, the Hall conductivity vanishes in all parameter regimes due to the TRS. 

So far we have analysed the phase diagram in the $u<0$, $c>0$ region. As shown in Appendix~\ref{app:bandProp}, the Hamiltonian possesses certain symmetries which relate phases in the remaining regions of the phase diagram to those in the $u<0$, $c>0$ region. Upon changing the sign of $u$, the $\mathbb{Z}_2$ invariant is preserved while the spin Hall conductivity changes sign. The spin Hall conductivity in $\tilde{P}_2$ is thus the negative of that in $\tilde{P}_1$. Upon changing the sign of $c$, both the $\mathbb{Z}_2$ invariant and the spin Hall conductivity remain unchanged.

\subsection{Slow quenches with preserved TRS}

Now we turn to the discussion of quenches. We studied the response of the system undergoing a slow quench of the parameters of the Hamiltonian between different topological regimes, indicated by the dashed lines in Fig.~\ref{fig:PhaseDiagram} (a). The parameter $u$ is changed smoothly as $u(t)=u_0+(u_1-u_0) \sin^2(\frac{\pi}{2}t/\tau_u)$ for $t\in[0,\tau_u]$. For times $t>\tau_u$, $u$  has a constant value $u_1$. During the quench the Hamiltonian stays time-reversal symmetric.

\begin{figure*}
	\centering
	{\includegraphics[width=145pt]{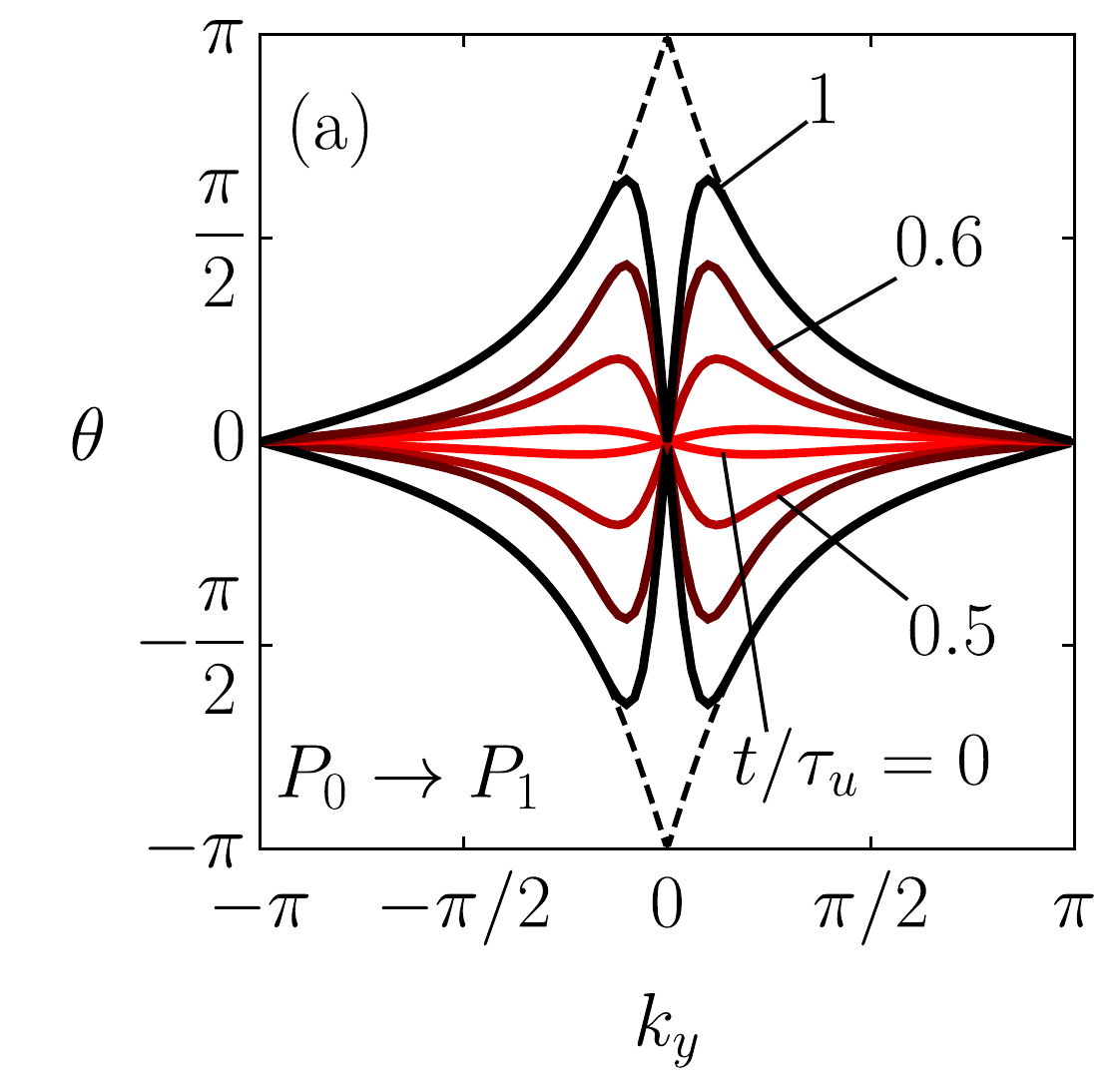}\includegraphics[width=145pt]{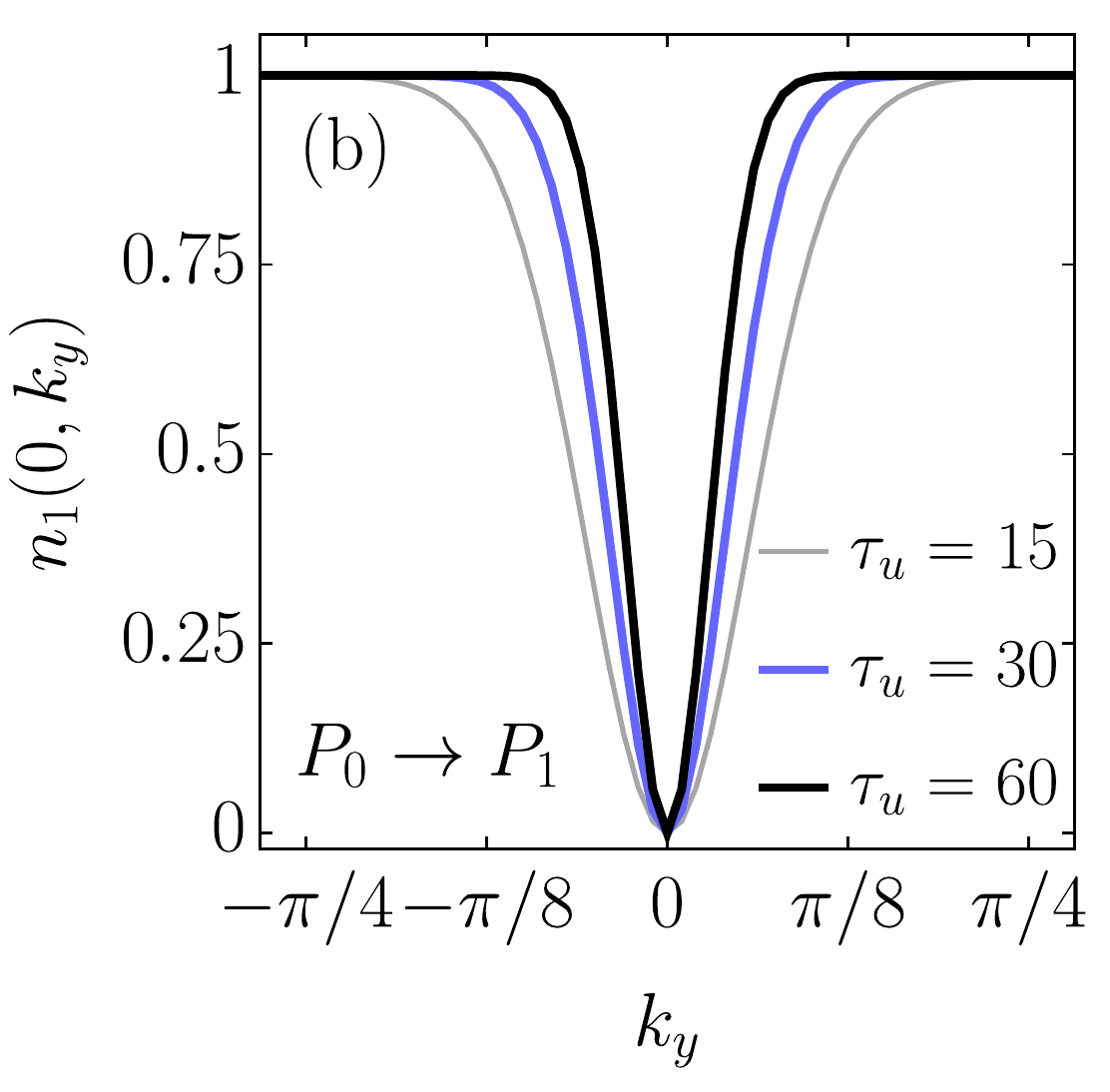}\includegraphics[width=145pt]{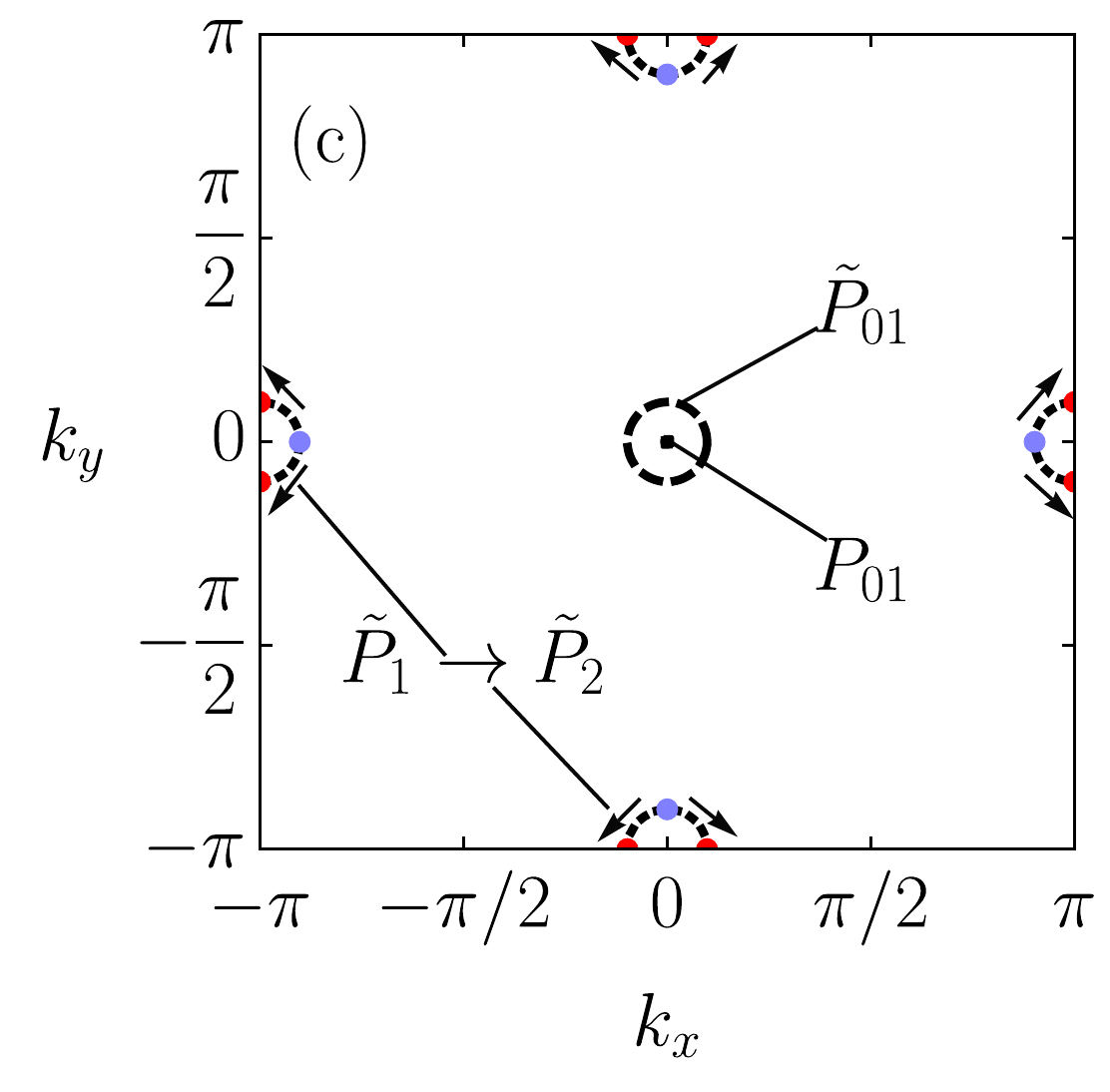}}
	
	\caption{ (a) Wannier centre flows of the ground state at $P_1$ (black, dashed) and of non-equilibrium states at various times during the quench from $P_0$ to $P_1$ with  $\tau_u=15$: $t/\tau_u=0$ (red), $t/\tau_u=0.5$ (dark red), $t/\tau_u=0.6$ (darker red), and after the quench at $t/\tau_u=1$ (black). (b) Population of the first energy level after the $P_0\to P_1$ quench with $\tau_u =15$ (grey, thin), $\tau_u =30$ (blue) and  $\tau_u =60$ (black). (c) Band gap closing in $k$ space. For $P_0 \to P_1$ and $\tilde{P}_0 \to \tilde{P}_1$ quenches the band gap closes at $P_{01}$ and $\tilde{P}_{01}$, respectively. However, for $\tilde{P}_1 \to \tilde{P}_2$ quench, it closes gradually starting at blue points and moving to red ones, as indicated with arrows.}
	\label{fig:QuenchTRSBand}
\end{figure*}

We first discuss the $c=0$ case. In Fig.~\ref{fig:QuenchTRSBand} (a) the Wannier centre flows are shown at different times during the quench $P_0 \to P_1$ with $\tau_u=15$. The system starts in the trivial phase with the shape of the Wannier centre flow as in Fig.~\ref{fig:PhaseDiagram} (b) (red line). With progressing time, the Wannier centre flow evolves into the diamond shape characteristic of the ground state at $P_1$ (black, dashed) for $|k_y|$ larger than a certain $k_0$, but deviates from that for $|k_y|<k_0$. 
Since the Wannier centre flow vanishes at $k_y = 0$ for all times, the $\mathbb{Z}_2$ invariant remains unchanged (see Appendix~\ref{app:proofs}).

The shape of the Wannier centre flow can be related to the band occupancy shown in Fig.~\ref{fig:QuenchTRSBand} (b). The population of the (doubly degenerate) lowest energy level $n_1(0,k_y)$ after the quench is shown for several quench times $\tau_u$. 
The population of the lowest energy level drops at small $|\mathbf{k}|<k_0$ and vanishes at the $\Gamma$ point, where the band gap closes during the quench. The final occupancies can also be derived analytically using the Landau-Zener formula \cite{ZenerLandau} which gives $n_1(\mathbf{k})=1-\exp[-\pi k^2/v_u]$, $v_{u}=|\left.\frac{\mathrm{d}u}{\mathrm{d}t}\right|_{t=\frac{\tau_u}{2}}|=\pi\left|u_{1}-u_{0}\right|/2\tau_{u}$, and consequently the delimiting $k_0\sim \sqrt{1/\tau_u}$ (see Appendix~\ref{app:bandProp}). The distance from the $\Gamma$ point thus determines how close the final state is to the ground state of the final Hamiltonian (at the corresponding $\mathbf{k}$), which explains the shape of the Wannier flows.

Fig. \ref{fig:QuenchTRSBand} (c) shows the $k$ points at which the band gap closes during the considered quenches. In contrast to the $P_0 \to P_1$ quench where the band gap closes at $P_{01}$ at an isolated point, for the $\tilde{P}_0\to \tilde{P}_1$ quench, the band gap closes at parameters $\tilde{P}_{01}$  on a circle around the $\Gamma$ point. The quench $\tilde{P}_1\to \tilde{P}_2$ is distinct from the former two, as parameters of the Hamiltonian enter the conducting region (see Fig.~\ref{fig:PhaseDiagram} (a)) and the band gap remains closed in a range of parameter values near $u=0$.
On entering the conducting region at $\tilde{P}_{12}$, the band gap closes at four points (blue points in Fig.~\ref{fig:QuenchTRSBand} (c)) which with increasing $u$ split into eight points that with the progressing quench move along circles centred at momenta $(\pm \pi,0)$ and $(0,\pm \pi)$.  When they reach  the Brillouin zone boundary (red points in Fig.~\ref{fig:QuenchTRSBand} (c)), the band gap opens and the insulating topological regime is reached. 

After the quench with $0<c\ll 1$, the electrons with momenta close to those on a circle where the band gap closes during the quench are excited to the lower conduction band with probability $\exp[-\pi (|\mathbf{k}-\mathbf{k}_c|-c)^2/v_u]$, where $\mathbf{k}_c$ is the centre of the circle. At momenta on the circle the population of the conduction band is $1$. The Wannier centre flow is deformed in such a way that the $\mathbb{Z}_2$ invariant is ill-defined (see Appendix~\ref{app:proofs}).

The total number of excitations is $N_{exc}=|u_1-u_0|/8\pi\tau_u$ for systems with $c=0$ and $N_{\mathrm{exc}}=c\sqrt{|u_{1}-u_{0}|/8\pi\tau_{u}}$ for systems with $0<c\ll 1$, i.e., in systems with zero and non-zero $c$ it scales differently with $\tau_u$. As $u$ is a linear function of time near the gap closing at $t=\tau_u/2$, we can compare these results to the predictions of the Kibble-Zurek scaling for linear quenches with a fixed rate, $N_{exc}\propto \tau_u^{-\nu d / (\nu z +1)}$. Here, $d$ is the spatial dimension while $\nu$ and $z$ are the correlation length and the dynamical critical exponent, respectively, associated with the critical point across which the system is quenched. Following Ref.~\onlinecite{WeiChen16} (see Appendix~\ref{app:bandProp}), we obtained the critical exponents of the BHZ model. These give the same behaviour as found from the Landau-Zener formula and read $\nu=1$ and $z=1$ for $c=0$ and $\nu=1/2$ and $z=2$ for $0<c\ll 1$. Systems with zero and non-zero $c$ thus belong to different universality classes.

\begin{figure*}
	\centering
	{\includegraphics[width=145pt]{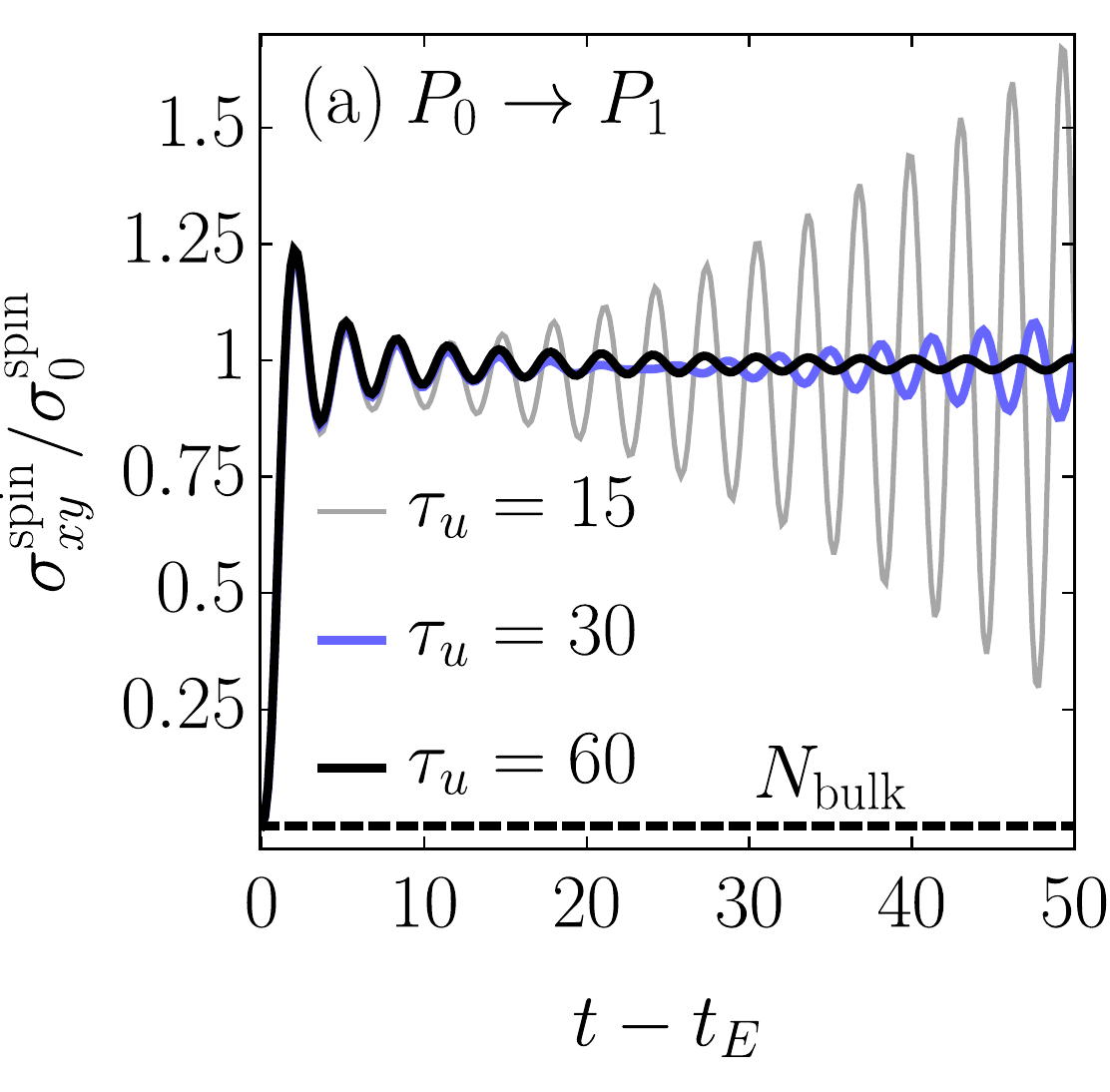}\includegraphics[width=145pt]{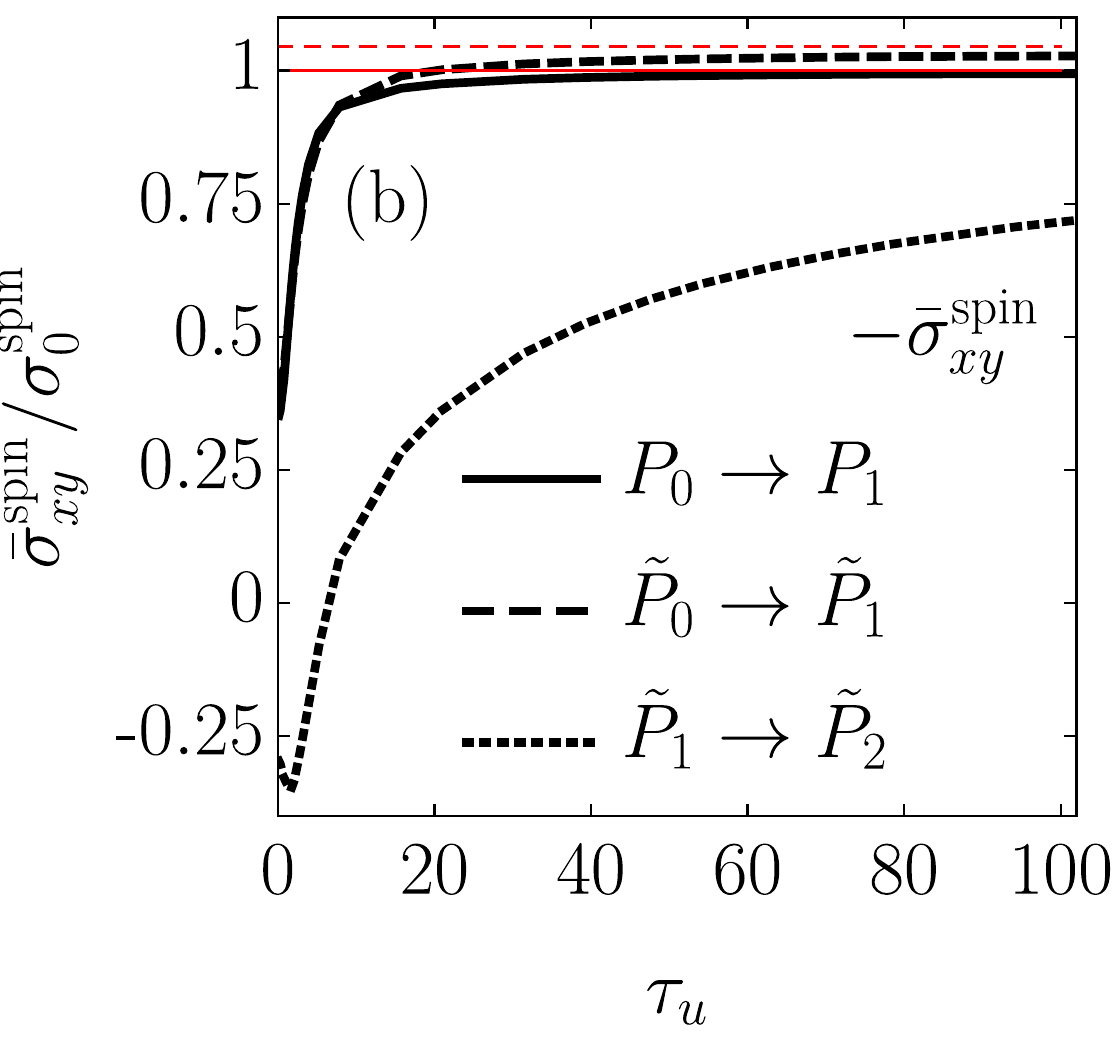}\includegraphics[width=145pt]{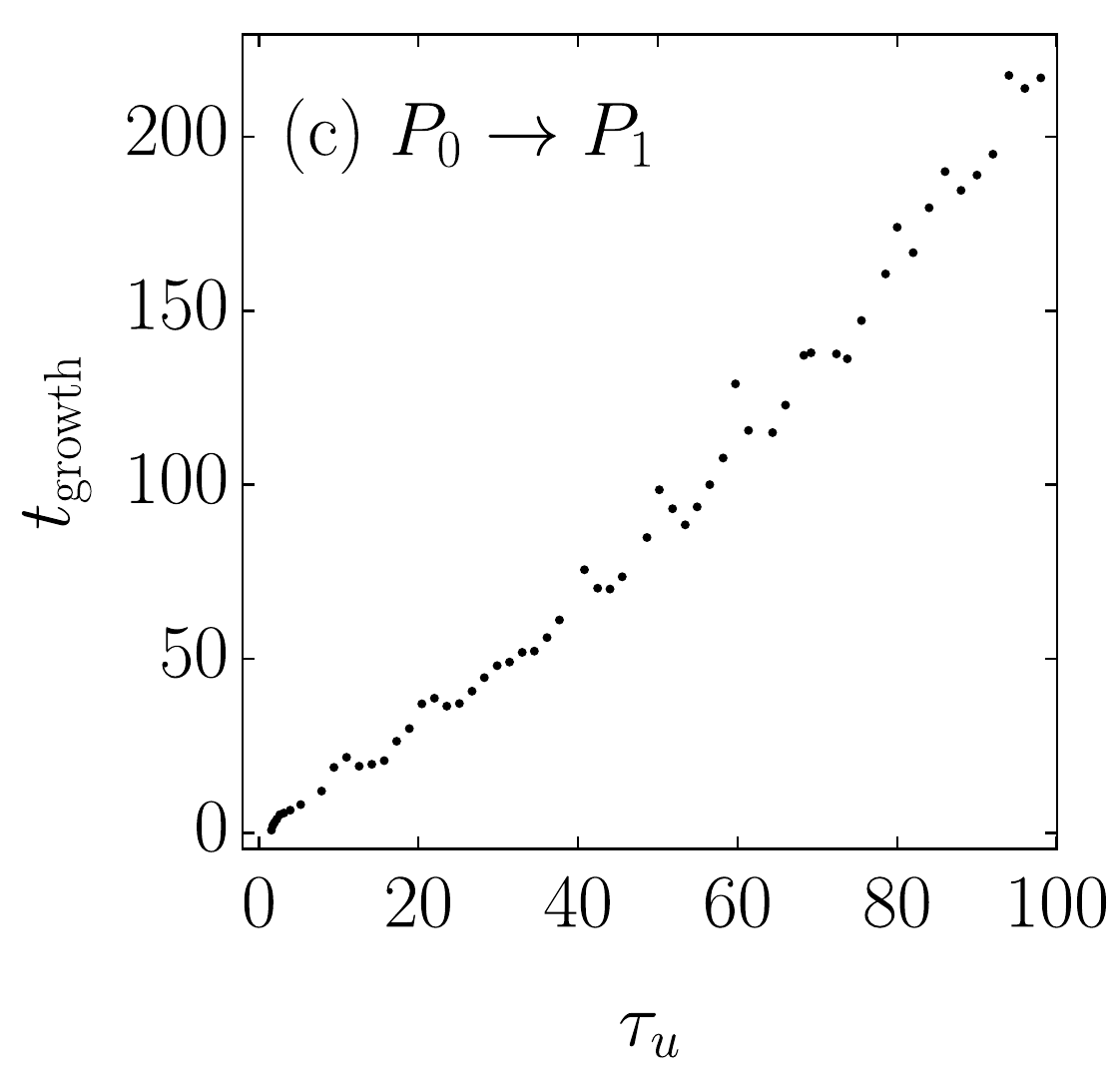}}
	
	\caption{(a) Topological invariant (black, dashed) and the spin Hall conductivity of the non-equilibrium states, resulting from the quench from $P_0$ to $P_1$, with $\tau_u=15$ (grey, thin), $\tau_u=30$ (blue) and $\tau_u=60$ (black). (b) Time-averaged spin Hall conductivity after the $P_0 \to P_1$ quench (black, solid) converges for $\tau_u\to\infty$ to the ground-state value of $\hat{H}(P_1)$ (red) and for systems after the $\tilde{P}_0 \to \tilde{P}_1$ (black, dashed) to the ground-state value of $\hat{H}(\tilde{P}_1)$ (red, dashed). Time-averaged spin Hall conductivity after the $\tilde{P}_1 \to \tilde{P}_2$ quench (black, dotted) is multiplied by a factor of $-1$ and converges to the ground-state value of $\hat{H}(\tilde{P_2})$. (c) Time at which the amplitude of oscillations in the spin Hall response after the quench $P_0 \to P_1$ increases for $10\%$  above the minimal amplitude.}
	\label{fig:QuenchTRSCond}
\end{figure*}

Spin Hall conductivities, evaluated as the electric field is turned on after the quench, are for several quench durations shown in Fig.~\ref{fig:QuenchTRSCond} (a). As for systems in the ground state, the spin Hall conductivity first experiences transient behaviour and then oscillates around a non-zero value $\bar{\sigma}_{xy}^\mathrm{spin}$, with the frequency equal to the band gap of the final Hamiltonian. As seen in the plot and as discussed in more detail below, the oscillations become small for slow enough quenches and $\bar{\sigma}_{xy}^\mathrm{spin}$ approaches the value characteristic of the final Hamiltonian. This generalizes the corresponding findings in Chern insulators in Ref.~\onlinecite{zoller}.

The dependence of $\bar{\sigma}_{xy}^\mathrm{spin}$  on the quench duration $\tau_u$ is presented in Fig.~\ref{fig:QuenchTRSCond} (b) for several quenches. The deviations from the ground-state values vanish for quenches slow enough. In order to understand the observed behaviour, we  evaluated the spin Hall conductivity using the time-dependent perturbation theory (following the discussion for Chern insulators in Ref. \onlinecite{Caio2016}, see also Appendix~\ref{app:perturb}). We obtain the following analytical formula for the time-averaged spin Hall conductivity at large times:
\begin{widetext}
\begin{equation}
\bar{\sigma}^\mathrm{spin}_{xy}=\frac{e}{(2\pi)^2}\sum_{n=1}^{2\,N_F}\intop\mathrm{d}\mathbf{k}\,n_n(\mathbf{k})\Omega^\mathrm{spin}_{n}(\mathbf{k}),\label{eq:SxyCond}
\end{equation}
\begin{equation}
\Omega^\mathrm{spin}_{n}(\mathbf{k})=-2\,\mathrm{Im}\sum_{m=1\atop m\neq n}^{2\,N_F}\frac{\langle u_n(\mathbf{k})|\frac{1}{2}\hat{s}_z\partial_{k_y}\hat{H}(\mathbf{k})|u_m(\mathbf{k})\rangle\langle u_m(\mathbf{k})|\partial_{k_x}\hat{H}(\mathbf{k})|u_n(\mathbf{k})\rangle}{(E_n(\mathbf{k})-E_m(\mathbf{k}))^2},\label{eq:SpinBerryCurv}
\end{equation}
\end{widetext}
where $n_n(\mathbf{k})$ is the occupation of the $n$-th energy band at $\mathbf{k}$ with eigenenergy $E_n(\mathbf{k})$. The spin Hall conductivity is expressed as an integral of the spin Berry curvature $\Omega^\mathrm{spin}_{n}(\mathbf{k})$ \cite{SpinBerryCurv} weighted by the band occupancy. The spin Berry curvature of conduction bands has the opposite sign to that of the valence bands. Therefore, the excitations above the ground state diminish the time-averaged spin Hall conductivity, which explains the dependence seen in Fig.~\ref{fig:QuenchTRSCond} (b). For slow quenches excitations occur in a small region in $k$ space and thus contribute little to the integral in Eq.~\eqref{eq:SxyCond}. Therefore, the time-averaged conductivity converges for long $\tau_u$ to the ground-state one. For systems with $c=0$ the time-averaged value for slow quenches can be further simplified to $\bar{\sigma}_{xy}^{\mathrm{spin}} \approx \frac{e}{2\pi}(1-|u_1-u_0|/4 \tau_u)$, where we used the Landau-Zener formula for the energy band occupancy and approximated the spin Berry curvature with its value at the $\Gamma$ point (similar was done in Ref.~\onlinecite{Unal16}). A similar calculation for systems with $c\neq 0$ shows that the deviation of the post-quench spin Hall response from the ground state value diminishes for long quenches as $\delta \sigma_{xy}^\mathrm{spin}\propto 1/\sqrt{\tau_u}$. For slow enough quenches $\delta \sigma_{xy}^\mathrm{spin}$ is proportional to the total number of excitations and hence obeys the Kibble-Zurek scaling.

The formula Eq.~\eqref{eq:SpinBerryCurv} is also useful for discussion of the different magnitudes of the deviations from ground-state values for different quench protocols as seen in Fig.~\ref{fig:QuenchTRSCond} (b) (straight lines denote ground-state values). 
Namely, after the $\tilde{P}_1\to \tilde{P}_2$ quench, the response deviates from the ground-state result much more than the responses after the other two quenches. This is due to two facts. First, the number of produced excitations is larger and, second, the excitations occur at momenta where the spin Berry curvature is large (see Appendix~\ref{app:bandProp}). More precisely, the number of excited electrons is approximately two times larger for $\tilde{P}_1\to \tilde{P}_2$  than for the $\tilde{P}_0\to \tilde{P}_1$ quench. The number is roughly given by the length of the circles in Fig.~\ref{fig:PhaseDiagram} (c). During the $\tilde{P}_1\to \tilde{P}_2$ quench, the band gap closing points cover two full circles. Second, the value of the spin Berry curvature where those excitations occur is larger for the former quench protocol. For the $P_0\to P_1$ and the $\tilde{P}_0\to \tilde{P}_1$ protocols, the spin Berry curvatures are small in the region with excited electrons, hence the deviations from the ground-state value of the spin Hall conductivity are smallest there. 

We now consider the oscillations around the time-averaged value. For short times oscillations diminish and for later times they start to grow quadratically (see Appendix~\ref{app:perturb}). The time $t_\mathrm{growth}$ after which the amplitude of oscillations starts to grow increases with the duration of the quench $\tau_u$. In order to give a somewhat more quantitative estimate of the behaviour, we defined $t_\mathrm{growth}$ as the time after which the amplitude of the oscillations increases by 10\% above the minimal amplitude found for a given $\tau_u$,  Fig.~\ref{fig:QuenchTRSCond} (c). Note that $t_\mathrm{growth}$ is roughly linear in $\tau_u$, which means that for slow quenches there is a long time window where these oscillations are not important. In Appendix~\ref{app:perturb}, we show that the growth of the oscillations occurs due to the non-zero off-diagonal elements of the density matrix in the basis of the eigenstates of the final Hamiltonian. Actually, often in evaluations of the Hall conductivity \cite{Wang16,Rigol2008,LevRigol2016} only the diagonal parts of the density matrix are retained, which is supported by the argument that a measurement of the Hall conductance unavoidably introduces decoherence and collapses the quenched state to a state represented by a diagonal ensemble. 

\subsection{Slow quenches with symmetry breaking}

\begin{figure*}
	\centerline{\includegraphics[width=145pt]{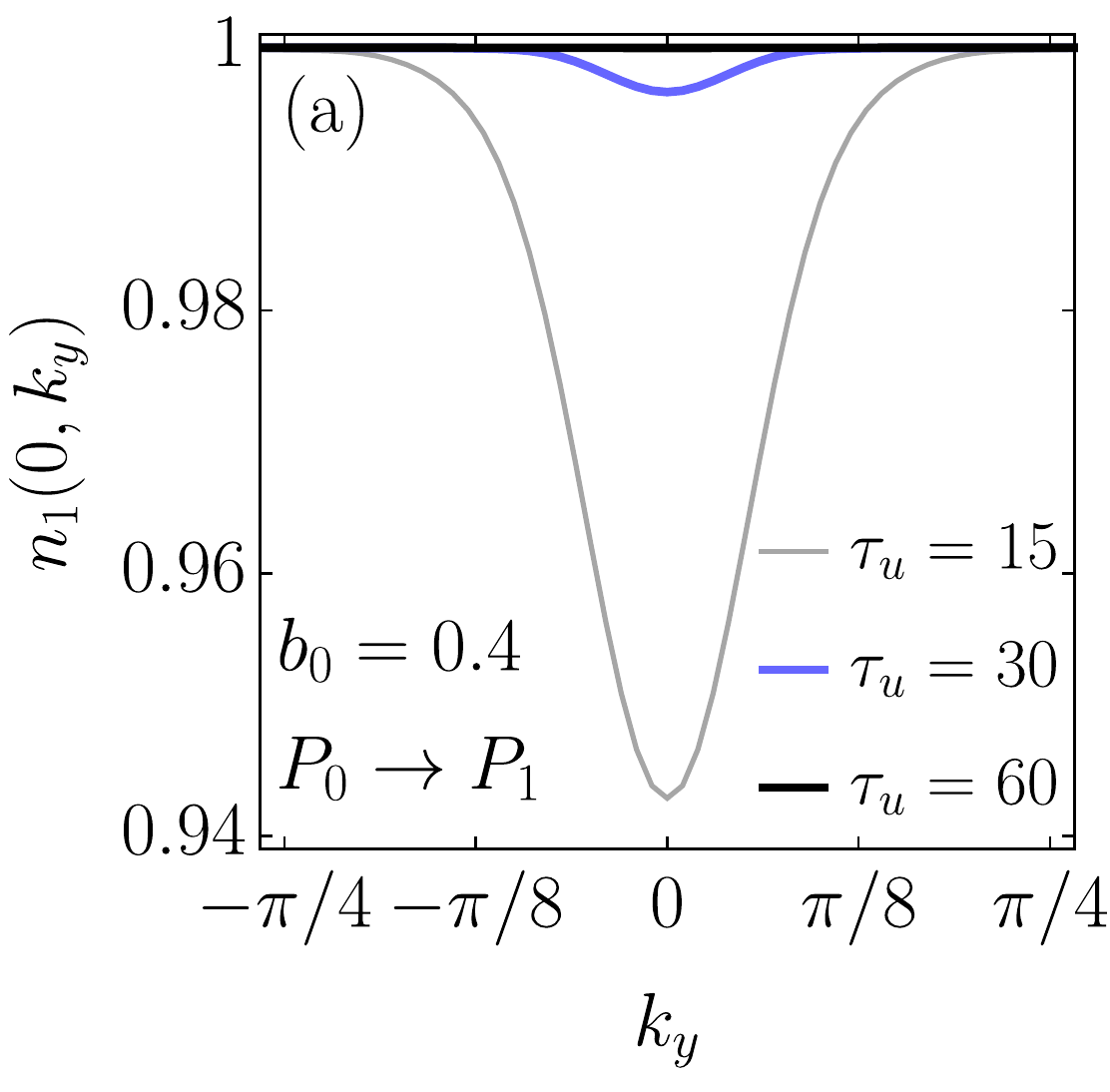}\includegraphics[width=145pt]{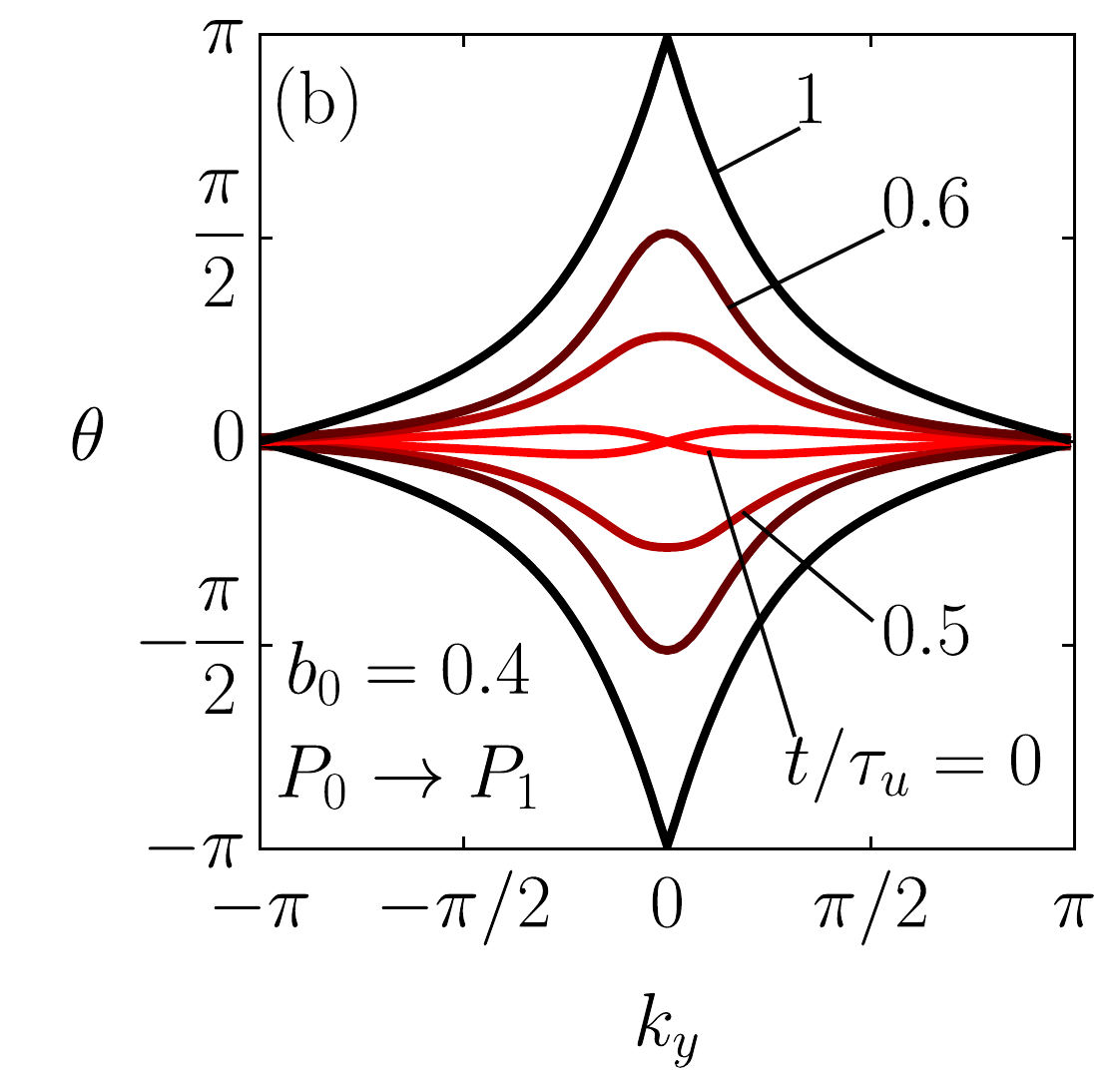}\includegraphics[width=145pt]{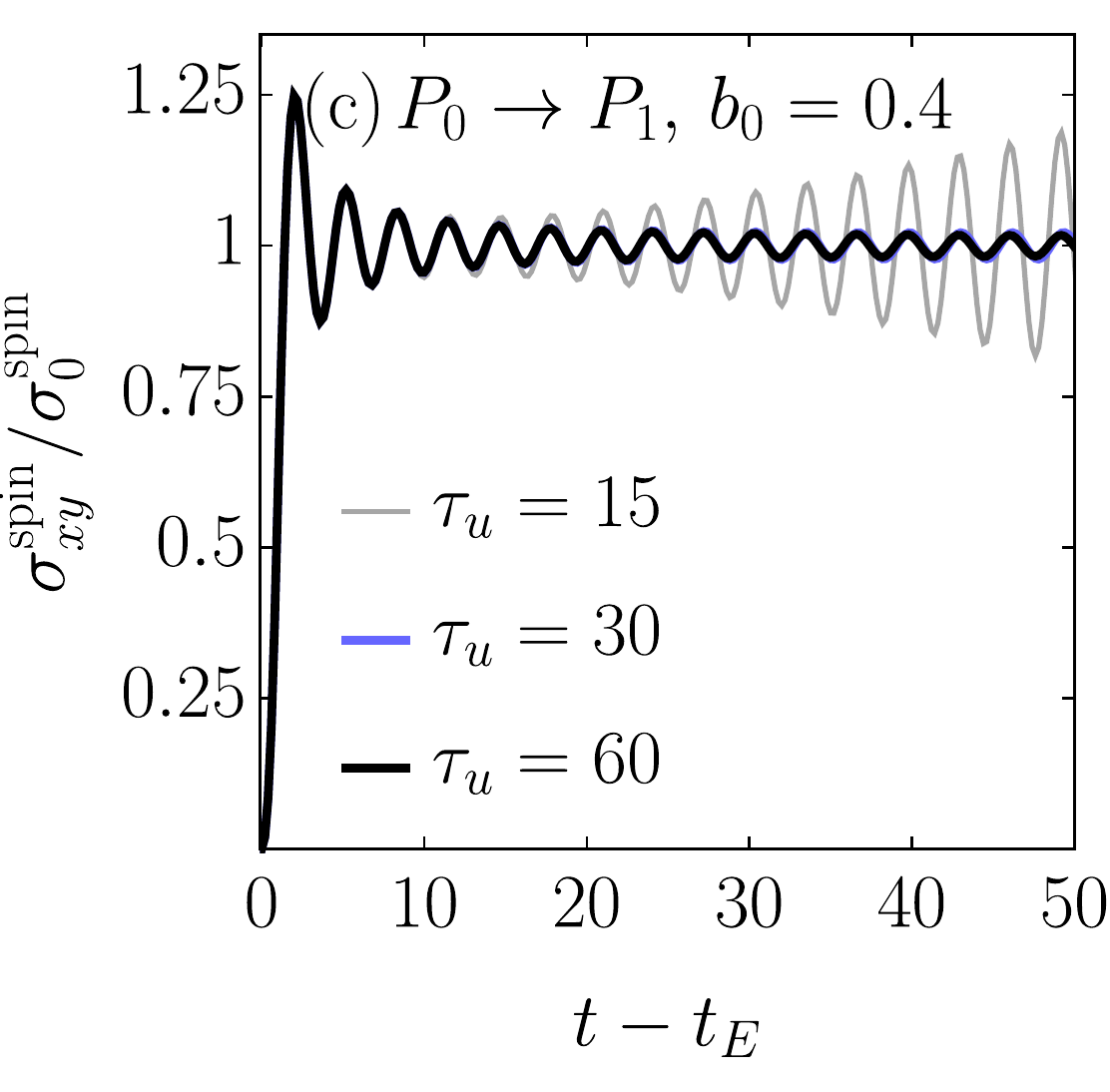}}
	\caption{Properties of the system quenched from $P_0$ to $P_1$ with a TRS breaking term of amplitude $b_0=0.4$. (a) Population of the first energy level $n_1(0,k_y)$ after the quench for $\tau_u =15$ (grey, thin), $\tau_u =30$ (blue) and $\tau_u =60$ (black). (b) Wannier centre flows of the non-equilibrium state at different times during the quench with $\tau_u=15$, $t/\tau_u=0$ (red), $t/\tau_u=0.5$ (dark red), $t/\tau_u=0.6$ (darker red), and $t/\tau_u=1$ (black). (c) Spin Hall conductivity of the systems after quenches with characteristic times $\tau_u=15$ (grey, thin), $\tau_u=30$ (blue) and $\tau_u=60$ (black). The latter two graphs overlap as the quench with such $\tau_u$ and $b_0$ is already adiabatic.}
	\label{fig:QuenchBreakTRS}
\end{figure*}

When an important symmetry of the Hamiltonian associated to a certain class of topological insulators is broken during a quench, different topological ground states can become adiabatically connected \cite{madzari}, i.e., the band gap can remain open everywhere during the quench. The topological invariant becomes ill-defined in this case. 
We study such processes by adding a convenient TRS breaking term $b \,\hat{s}_x \otimes \hat{\sigma}_x$ to the BHZ model. In parallel to changing the parameter $u$ as in the case of the TRS preserving quench, the amplitude $b$ is turned on during the quench as $b(t)=b_0 \sin^2(\pi t/\tau_u)$. In this way the Hamiltonian has the TRS before and after the quench but for $0<t<\tau_u$ the symmetry is broken and the band gap remains open. When the quench is done slowly enough compared to the inverse of the minimal band gap during the quench, there are almost no excitations to conduction bands. This can be seen in Fig.~\ref{fig:QuenchBreakTRS} (a) where the population of the first energy band after the $P_0\to P_1$ quench is shown for various quench times. In the adiabatic limit, the system ends up in the ground state of the final Hamiltonian and thus in the topological phase with $N_{\mathrm{bulk}}=1$.

Time-reversal properties of the system during the quench can be observed from the graphs of the Wannier centre flow in Fig.~\ref{fig:QuenchBreakTRS} (b). At $t/\tau_u=0$ and $t/\tau_u=1$ the Wannier centre flow has the typical form for the trivial and topological phase, respectively, while at $t/\tau_u=0.5$ and $t/\tau_u=0.6$ the system does not exhibit the TRS as can be seen by the absence of double degeneracy at $k_y=0$ and $k_y=\pi$.

After the quench, the electric field is turned on. At long times, the spin Hall response oscillates around a constant value with the frequency equal to the band gap of the final Hamiltonian, as shown in Fig.~\ref{fig:QuenchBreakTRS} (c). For a quench slower than the inverse of the minimal band gap during the quench, the system exhibits ground-state spin Hall response. It also coincides with the spin Hall response of the system after an infinitely slow symmetry preserving quench. For faster quenches, there are excitations present even in the symmetry breaking case (see Fig.~\ref{fig:QuenchBreakTRS} (a)) so the growth of oscillations and the deviation of the time-averaged value from the ground-state value can be observed. However, compared to the case of symmetry preserving quench, the oscillations are less prominent because of the smaller number of excitations.

\section{Conclusions}
\label{sec:concl}

In this paper we calculated the topological invariant and the transport properties of the time-reversal symmetric BHZ model undergoing a slow quench between different topological regimes. Similarly to the case of the Chern insulators discussed in the literature earlier, our results show that in the BHZ model that has besides the time-reversal symmetry also the inversion symmetry and conserves the spin projection $s_z$, such a quench preserves the bulk topological invariant Eq.~\eqref{Z2def}  (the conservation of this quantity is not a manifestation of the time-reversal symmetry that is dynamically broken \cite{McGinleyCooper} but rather due to the inversion symmetry). In a general case where $s_z$ is not conserved, the $\mathbb{Z}_2$ invariant becomes ill-defined after the quench. 

The spin Hall response  for slow enough quenches approaches that of the ground state of the final Hamiltonian. The transport properties that are given as an integral over the Brillouin zone can universally be expected to be close to those of the final Hamiltonian ground state as the quench is adiabatic for all states except for those in a small region in the momentum space. Hence, for the cases where the bulk invariant is conserved, the loss of correspondence of the bulk invariant and the bulk transport properties is expected. It would be interesting to explore the behaviour of bulk invariants during quenches in other systems, too.

We also considered quenches during which the TRS of the time-dependent Hamiltonian is broken, which allows the adiabatic connection between different regimes and hence restores the correspondence between the bulk invariant and the transport.  

It would be interesting to investigate the dynamics of time-reversal symmetric systems also experimentally with HgTe/CdTe and InAs/GaSb/AlSb Type-II semiconductor quantum wells where the quench could perhaps be performed by varying the inversion breaking electric potential in the $z$-direction, which can be tuned by a top gate in experiments \cite{RashbaClanek,TMOlayer,MagTopIns}.  Alternatively, time-dependent Hamiltonians can be also realized in ultracold atoms~\cite{Bloch2013,Ketterle2013,Esslinger2014,Bloch2015,Flascher2016}.

\acknowledgements

We acknowledge useful discussions with B. Donvil, L. Vidmar, and I. \v{Z}uti\'c. The work was supported by the Slovenian Research Agency under contract no. P1-0044.

\appendix

\section{BAND PROPERTIES AT AND AFTER BAND GAP CLOSING}
\label{app:bandProp}
\subsection{Band gap closings and post-quench occupancies}

\begin{figure*}
	\centering \includegraphics[width=440pt]{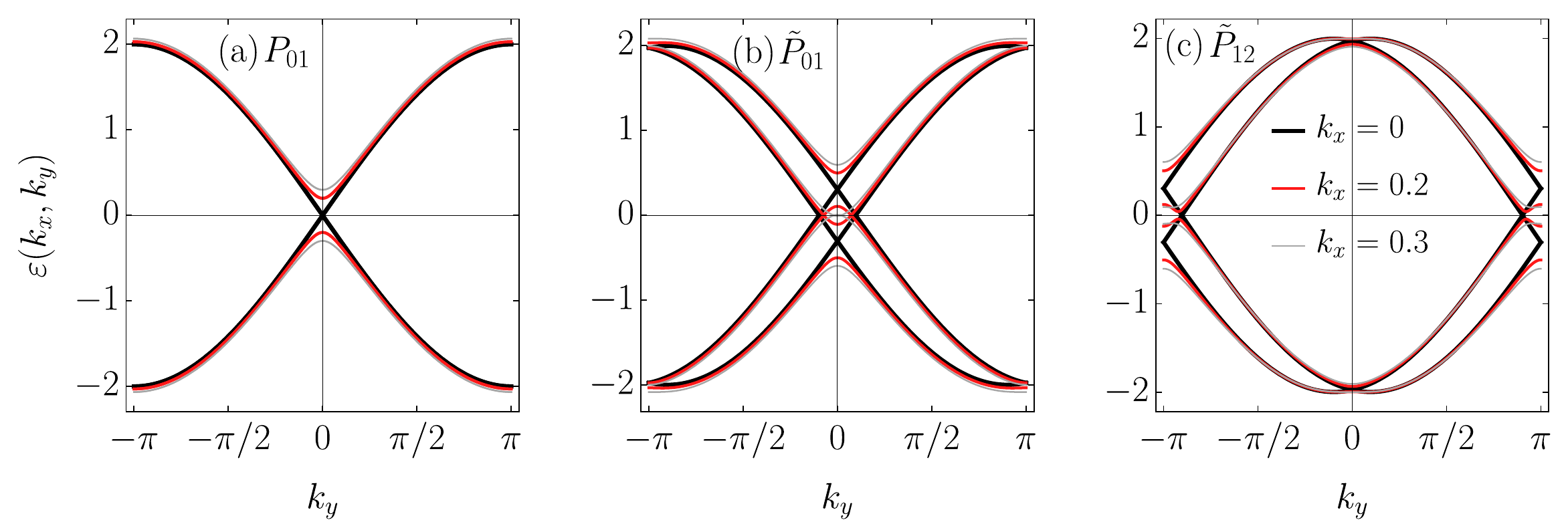}	
	\caption{ Band dispersions at (a) $P_{01}$, (b) $\tilde{P}_{01}$ and (c) $\tilde{P}_{12}$
		for $k_{x}=0$ (black), $k_{x}=0.2$ (blue) and $k_{x}=0.3$ (red).}
	\label{fig:BandClosing} 
\end{figure*}

Let us introduce $u_{c}$ and $\mathbf{k}_{c}$ which are the values
of $u$ and $\mathbf{k}$, respectively, where the band gap closes
along the $c=0$ line of the phase diagram in Fig.~\ref{fig:PhaseDiagram}~(a).
For quenches discussed in this paper the relevant band closings are
those at $u_{c}=-2$ (the gap closes at $\mathbf{k}_{c}=\left(0,0\right)$) and at $u_{c}=0$
(the gap closes at $\mathbf{k}_{c}=(\pi,0)$ and at $\mathbf{k}_{c}=\left(0,\pi\right)$).
For system with $0<c\ll1$ ($c=0.3$ is small enough), the band gap closes at momenta
close to those at $c=0$.
Expanding the Hamiltonian (\ref{eq:H}) to the first order in the
deviation of the momentum from $\mathbf{k}_{c}$ we obtain band dispersions
$\pm\sqrt{(q-c)^{2}+\delta u^{2}}$ and $\pm\sqrt{(q+c)^{2}+\delta u^{2}}$,
where $q=\left|\mathbf{\mathbf{k}}-\mathbf{k}_{c}\right|\ll\pi$ and
$\delta u=u-u_{c}$. For $c=0$ the band gap between two spin degenerate
valence bands and two spin degenerate conduction bands closes at $\mathbf{k}_{c}$
with linear dispersion $\pm q$ while for $0<c\ll1$ the band gap
between the upper valence band and the lower conduction band closes
on a circle with radius $c$ around $\mathbf{k}_{c}$, again with
linear dispersion $\pm\left|q-c\right|$. Fig.~\ref{fig:BandClosing}
shows cross-sections of band dispersions for different $k_{x}$ at
parameters $P$ for which the band gap closes during quenches discussed
in this paper. Note that in the case of the $\tilde{P}_{1}\to \tilde{P}_{2}$
quench there is a semimetal region of a finite width around $u=0$
in the phase diagram. While the band gap closes on a circle as discussed
above, for different points along the circle this happens at different
values of $u$ inside the semimetal region. For example, for the particular
value of $u$ corresponding to Fig.~\ref{fig:BandClosing}~(c) the
gap is closed only at $k_{x}=0$.

\begin{figure*}
	\includegraphics[width=440pt]{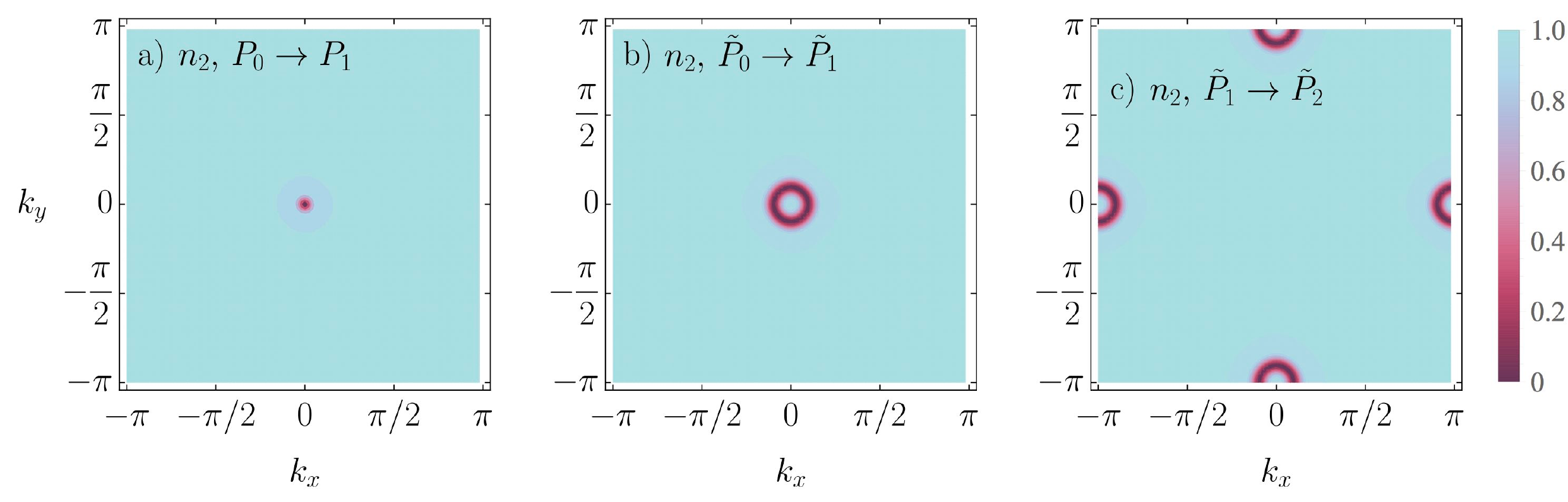}
	
	\caption{Occupancy of the upper valence band of the BHZ model quenched between
		(a) $P_{0}\to P_{1}$, (b) $\tilde{P}_{0}\to\tilde{P}_{1}$, and (c)
		$\tilde{P}_{1}\to\tilde{P_{2}}$, for $\tau_{u}=60$.}
	\label{fig:Occupancy2D} 
\end{figure*}

Near the band gap closing, i.e., for $\left|\delta u\right|\ll1$ and
$\left|q-c\right|\ll1$, the low-energy physics is described by the
Landau-Zener Hamiltonian. The probability for the transition from
the valence to the conduction band is given by the Landau-Zener
formula $e^{-\pi(q-c)^{2}/v_{u}}$, where $v_{u}=|\left.\frac{\mathrm{d}u}{\mathrm{d}t}\right|_{u=u_{c}}|$ \cite{ZenerLandau}.
In our case $v_{u}=\pi\left|u_{1}-u_{0}\right|/2\tau_{u}$. Excitations
to the conduction band occur at momenta where $\left|q-c\right|\lesssim\sqrt{v_{u}}$,
i.e. on a disk of radius $\sim\sqrt{v_{u}}$ for $c=0$ and, provided
$\tau_{u}$ is long enough so that $\sqrt{v_{u}}\ll c$, on a ring
of radius $c$ and of width $\sim\sqrt{v_{u}}$ for $0<c\ll1$. Fig.~\ref{fig:Occupancy2D}
shows occupations of the upper valence band after the quenches considered
in this paper. As $v_{u}\propto1/\tau_{u}$, the total
number of electrons excited to the conduction band around one gap closing $N_{\textrm{exc}}$
is proportional to $1/\tau_{u}$ for $c=0$, while for $0<c\ll1$
it is proportional to $c/\sqrt{\tau_{u}}$. A more detailed calculation
yields $N_{\mathrm{exc}}=|u_{1}-u_{0}|/8\pi\tau_{u}$
and $N_{\mathrm{exc}}=c\sqrt{|u_{1}-u_{0}|/8\pi\tau_{u}}$,
respectively. Note that for the quench $\tilde{P}_1\to \tilde{P}_2$ the gap closes on two separate circles and for the quench $P_0\to P_1$ between two pairs of bands, so the number of excitations is twice as high.

\subsection{Spin Berry curvature and post-quench response}

An intriguing observation made in Fig.~\ref{fig:QuenchTRSCond}~(b)
is that after the $\tilde{P}_{1}\to\tilde{P}_{2}$ quench, the time-averaged
spin Hall conductivity $\bar{\sigma}_{xy}^{\mathrm{spin}}$ deviates
from the ground state value of the final Hamiltonian for an order
of magnitude more than after the $\tilde{P}_0\to\tilde{P}_1$ quench. Here we provide
a detailed explanation of this puzzling behaviour.

\begin{figure*}
	\centering \includegraphics[width=440pt]{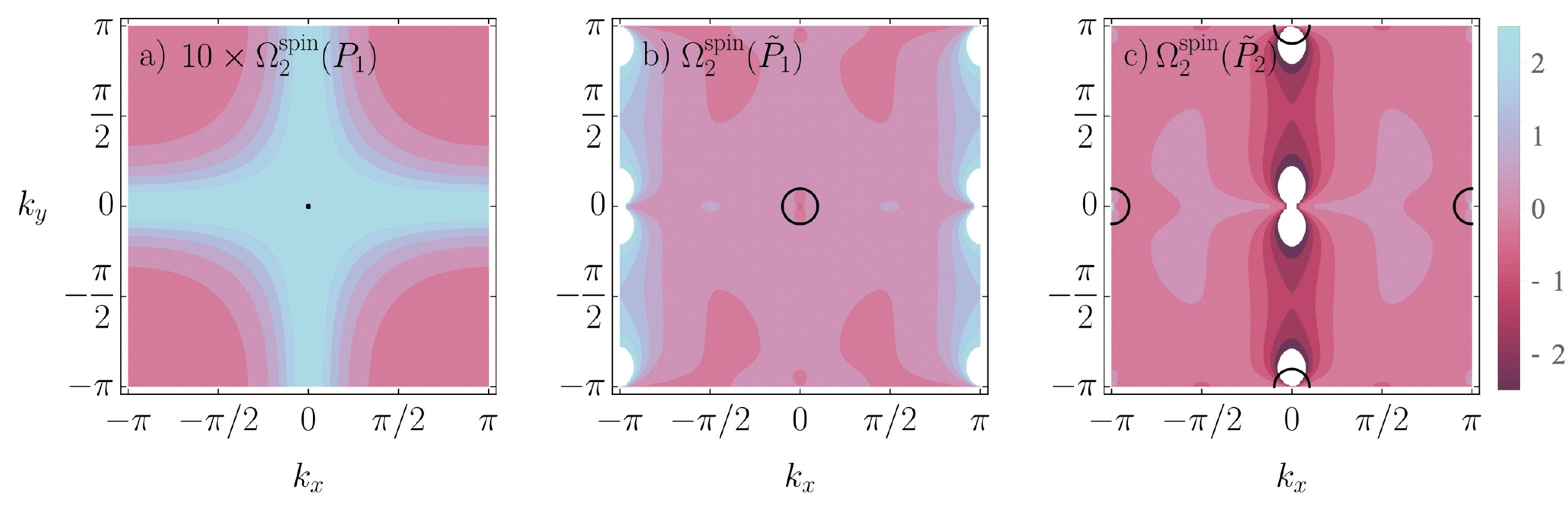}
	
	\caption{Spin Berry curvature of the upper valence band of the BHZ model at
		parameters (a) $P_{1}$, (b) $\tilde{P}_{1}$, and (c) $\tilde{P}_{2}$.
		The value of the spin Berry curvature at $P_{1}$ is multiplied by
		$10$. Black lines denote momenta at which the band gap closes during
		quenches (a) $P_{0}\to P_{1}$, (b) $\tilde{P}_{0}\to\tilde{P}_{1}$,
		and (c) $\tilde{P}_{1}\to\tilde{P}_{2}$. White regions correspond to the values out of colour code scale.}
	\label{fig:SpinBerryCurv} 
\end{figure*}

For $\tau_{u}$ large enough, the deviation of the post-quench time-averaged
spin Hall conductivity from its value in the ground state of the post-quench
Hamiltonian can be expressed as
\begin{equation}
\delta\bar{\sigma}_{xy}^{\mathrm{spin}}\approx-2e\sum_{\kappa}N_{\mathrm{exc},\kappa}\bar{\Omega}_{2,\kappa}^{\mathrm{spin}},
\end{equation}
where $\kappa$ runs over the band gap closings and $\bar{\Omega}_{2,\kappa}^{\mathrm{spin}}$ is the spin Berry curvature
of the upper valence band of the final Hamiltonian, averaged over the
momenta where the band gap closes for a particular $\kappa$. The prefactor $2$ comes from
the fact that both electrons excited to the conduction band as
well as holes left in the valence band contribute equally as the
spin Berry curvatures of those bands are opposite, $\Omega_{3}^{\textrm{spin }}\left(\mathbf{k}\right)=-\Omega_{2}^{\textrm{spin }}\left(\mathbf{k}\right).$
Fig.~\ref{fig:SpinBerryCurv} shows spin Berry curvatures of the
upper valence band of post-quench Hamiltonians discussed in this paper.
By considering Figs.~\ref{fig:Occupancy2D} and \ref{fig:SpinBerryCurv}
it is apparent that excitations to the conduction band after the $\tilde{P}_{1}\to\tilde{P}_{2}$
quench occur at momenta where the spin Berry curvature is for an order
of magnitude larger than the spin Berry curvature in the excitation
region after the $P_{0}\to P_{1}$ and $\tilde{P}_{0}\to\tilde{P}_{1}$
quenches.

As seen in Figs.~\ref{fig:SpinBerryCurv} (b) and \ref{fig:SpinBerryCurv} (c), the spin Berry curvature at $\tilde{P}_2$, when translated by $\pi$ in both $k_x$ and $k_y$ directions, is the negative of the spin Berry curvature at $\tilde{P}_1$. This is due to the fact that the Hamiltonian at $-u$ can be transformed into that at $u$: $\hat{s}_y\otimes \hat{\sigma}_x\hat{H}_{-u}(k_x+\pi,k_y+\pi)\hat{s}_y\otimes \hat{\sigma}_x=\hat{H}_u(k_x,k_y)$. A short calculation shows that the spin Hall conductivity of the ground state at $-u$ is of the opposite sign to the one at $u$ while the $\mathbb{Z}_2$ invariant is the same. Similarly, Hamiltonians at $-c$ and $c$ are related as $\hat{s}_z\hat{H}_{-c}(\mathbf{k})\hat{s}_z=\hat{H}_c(\mathbf{k})$. From this it can be shown that the spin Hall conductivity and the $\mathbb{Z}_2$ invariant are the same for the ground state at $c$ and $-c$. 

\subsection{Calculation of the critical exponents}

Let us choose a control parameter $\varepsilon$ such that the system undergoes a quantum phase transition at $\varepsilon=0$. A quantum phase transition is characterized by the divergence of both the characteristic length scale $\xi(\varepsilon)\propto\left|\varepsilon\right|^{-\nu}$ and characteristic time scale $\tau(\varepsilon)\propto \left|\varepsilon\right|^{-z\nu}$, $\nu$ being the correlation length and $z$ the dynamical critical exponent. According to the Kibble-Zurek argument, the scaling of the produced defect density, in our case excitations to conduction bands, depends on these critical exponents. In this section, we calculate the critical exponents of the BHZ model.

We extract the critical exponent $z\nu$ from the characteristic time scale, which is the inverse of the band gap. Noting that the spectrum near the gap closing is of the form $\pm\sqrt{(q\pm c)^{2}+\delta u^{2}}$, we find that the minimal gap vanishes as $\delta u^{z\nu}=\delta u$, yielding the critical exponent $z\nu=1$.

For the calculation of the correlation length critical exponent $\nu$ we follow Ref.~\onlinecite{WeiChen16}. Authors of Ref.~\onlinecite{WeiChen16} define the scaling function $F(\mathbf{k},\varepsilon)=(\mathbf{\hat{k}}_s\cdot \nabla_\mathbf{k})^2\mathrm{Pf}(m)$, where $\mathbf{\hat{k}}_s$ is the scaling direction and $m$ is the matrix of the time-reversal operator $\hat{\mathcal{T}}$ with elements $m_{\alpha\beta}(\mathbf{k},\varepsilon)=\langle u_\alpha(\mathbf{k},\varepsilon)|\hat{\mathcal{T}}|u_\beta(\mathbf{k},\varepsilon)\rangle$, $|u_\alpha(\mathbf{k},\varepsilon)\rangle$ being the occupied eigenstate $\alpha$ at momentum $\mathbf{k}$ and control parameter $\varepsilon$. The length scale is obtained from the scaling function $F(\mathbf{k},\varepsilon)$ at time-reversal symmetric momenta $\mathbf{k}_0$ as
\begin{equation}
\xi = \left|\frac{1}{\varepsilon}\frac{(\mathbf{\hat{k}}_s\cdot \nabla_\mathbf{k})^2F(\mathbf{k},\varepsilon)|_{\mathbf{k}=\mathbf{k}_0}}{\partial_\varepsilon F(\mathbf{k}_0,\varepsilon)}\right|^\frac{1}{2}.
\end{equation}
Using this approach, we obtain the critical exponent $\nu$ of the BHZ model: $\nu=1$ for $c=0$ and $\nu=1/2$ for $0<c\ll1$.

\section{Conservation of bulk invariants}
\label{app:proofs}
By definition, topological invariants do not change during adiabatic transformations of the Hamiltonian that respect the important symmetries. However, more general non-adiabatic transformations during which the band gap can even close, were found to preserve the Chern number (at least for two band systems)~\cite{PxPySuperFluid,Rigol2015},  too. Below we generalize the proof of the conservation of the Chern number to the case of multiple band systems and we discuss the conservation of the $\mathbb{Z}_2$ invariant of the BHZ model.

\subsection{Conservation of the Chern number}
\label{sec:ChernConser}
For a two-dimensional insulating non-interacting system with translational symmetry one can define the Chern number \cite{Berry} as
\begin{equation}
C=\frac{1}{2\pi}\sum_{n=1}^{N_F} \intop \mathrm{d}\mathbf{k}\, \Omega_n(\mathbf{k}),
\end{equation}
\begin{equation}
\Omega_n(\mathbf{k})=-i\partial_{k_x}\langle u_n(\mathbf{k})|\partial_{k_y}|u_n(\mathbf{k})\rangle+
i\partial_{k_y}\langle u_n(\mathbf{k})|\partial_{k_x}|u_n(\mathbf{k})\rangle,
\end{equation}
where $\Omega_n(\mathbf{k})$ is the Berry curvature of the $n$-th band. 

Let the system be in a state with the Chern number $C$ and the Berry curvature $\Omega_n(\mathbf{k})$ with $N_F$ occupied states $|u_n (\mathbf{k})\rangle$. We limit our discussion to transformations with translational symmetry described by a unitary operator $U(\mathbf{k})$, so each state is transformed as 
\begin{equation}
|u_n (\mathbf{k})\rangle \to |u'_n (\mathbf{k})\rangle =  U(\mathbf{k})|u_n (\mathbf{k})\rangle.\label{eq:evolvedState}
\end{equation}
The Berry curvature after the transformation is
\begin{align}
\begin{aligned}
\Omega'_n(\mathbf{k})=-i\partial_{k_x}\langle U(\mathbf{k}) u_n(\mathbf{k})|\partial_{k_y}|U(\mathbf{k}) u_n(\mathbf{k})\rangle\\
+
i\partial_{k_y}\langle U(\mathbf{k}) u_n(\mathbf{k})|\partial_{k_x}|U(\mathbf{k}) u_n(\mathbf{k})\rangle=\\
= 
\Omega_n(\mathbf{k}) -i\partial_{k_x}\langle u_n(\mathbf{k})|U(\mathbf{k})^{\dagger}[\partial_{k_y}U(\mathbf{k})]|u_n(\mathbf{k})\rangle\\+
i\partial_{k_y}\langle u_n(\mathbf{k})|U(\mathbf{k})^{\dagger}[\partial_{k_x}U(\mathbf{k})]|u_n(\mathbf{k})\rangle ,
\end{aligned}
\end{align}
where we used $\partial_{k_i}|U(\mathbf{k}) u_n(\mathbf{k})\rangle=[\partial_{k_i}U(\mathbf{k})]| u_n(\mathbf{k})\rangle+U(\mathbf{k})\partial_{k_i}| u_n(\mathbf{k})\rangle$. Note that expressions $\langle u_n(\mathbf{k})|\hat{O}(\mathbf{k})| u_n(\mathbf{k}) \rangle=\mathrm{Tr}[\rho_n(\mathbf{k})\hat{O}(\mathbf{k})]$ are smooth in $\mathbf{k}$ when $\hat{O}(\mathbf{k})$ and the density matrix $\rho_n(\mathbf{k})=|u_n(\mathbf{k})\rangle\langle u_n(\mathbf{k})|$ are smooth in $\mathbf{k}$. When $U(\mathbf{k})$ is smooth in $\mathbf{k}$,  which is true for time evolutions with Hamiltonians that are smooth in $\mathbf{k}$, the second
and the third term in $\Omega_n(\mathbf{k})$ are continuous functions of $\mathbf{k}$ and the Chern number can be written as
\begin{equation}
\begin{split}
C'= C+\frac{1}{2\pi}\sum_{n=1}^{N_F} \intop \mathrm{d}k_y  \intop \mathrm{d}k_x \,(-i)\partial_{k_x}\langle u_n(\mathbf{k})|U(\mathbf{k})^{\dagger}\\
\times[\partial_{k_y}U(\mathbf{k})]|u_n(\mathbf{k})\rangle
+\\
\frac{1}{2\pi}\sum_{n=1}^{N_F} \intop \mathrm{d}k_x \intop \mathrm{d}k_y \,i\partial_{k_y}\langle u_n(\mathbf{k})|U(\mathbf{k})^{\dagger}\\
\times[\partial_{k_x}U(\mathbf{k})]|u_n(\mathbf{k})\rangle.\label{eq:ChernConserv}
\end{split}
\end{equation}
Using the periodicity of $U(\mathbf{k})$ and $|u_n(\mathbf{k})\rangle$ over the Brillouin zone, the first integral over $k_x$ and the second integral over $k_y$ are zero (or, put differently, each component of the vector field $ \langle u_n(\mathbf{k})|U(\mathbf{k})^{\dagger}[\partial_{k_i} U(\mathbf{k})]|u_n(\mathbf{k})\rangle $ is smooth in $\mathbf{k}$, hence the Stokes theorem can be applied). Hence, we obtain $C'= C$, i.e. the Chern number is conserved under a unitary transformation.

To show that our proof reduces to the one for two-band systems done by D'Allesio and Rigol in Ref.~\onlinecite{Rigol2015}, we calculate the time derivative of the Chern number \eqref{eq:ChernConserv}. $U(\mathbf{k})$ now represents the time evolution operator and by taking into account that $\partial_tU^\dagger(\mathbf{k})[\partial_{k_i}U(\mathbf{k})]=-iU^\dagger(\mathbf{k})[\partial_{k_i}\hat{H}(\mathbf{k})]U(\mathbf{k})$, we get
\begin{equation}
\begin{split}
\partial_tC'=\frac{1}{2\pi}\sum_{n=1}^{N_F} \intop \mathrm{d}\mathbf{k} \,\Big(\partial_{k_y}\langle u'_n(\mathbf{k})|[\partial_{k_x}\hat{H}(\mathbf{k})]|u'_n(\mathbf{k})\rangle\\
-\partial_{k_x}\langle u'_n(\mathbf{k})|[\partial_{k_y}\hat{H}(\mathbf{k})]|u'_n(\mathbf{k})\rangle \Big).
\end{split}
\end{equation}
For a two-band Hamiltonian $\hat{H}(\mathbf{k})=-\frac{1}{2}\mathbf{B}(\mathbf{k})\cdot\hat{\boldsymbol{\sigma}}$ and the state expressed with the density matrix $\hat{\rho}(\mathbf{k})=\frac{1}{2}(\hat{\sigma}_0+\mathbf{S}(\mathbf{k})\cdot\hat{\boldsymbol{\sigma}})$, we obtain the result from Ref.~\onlinecite{Rigol2015}
\begin{equation}
\begin{split}
\partial_tC'=\frac{1}{4\pi}\intop \mathrm{d}\mathbf{k} \Big[ \partial_{k_x}(\mathbf{S}(\mathbf{k})\cdot \partial_{k_y}\mathbf{B}(\mathbf{k}))\\
-\partial_{k_y}(\mathbf{S}(\mathbf{k})\cdot \partial_{k_x}\mathbf{B}(\mathbf{k})) \Big],
\end{split}
\end{equation}
which vanishes after the application of the Stokes theorem due to the smoothness of $\mathbf{S}(\mathbf{k})\cdot \partial_{k_i}\mathbf{B}(\mathbf{k})$ at all times.
\subsection{Conservation of the $\mathbb{Z}_2$ invariant}

The quench dynamically breaks the TRS even if the time-dependent Hamiltonian has TRS \cite{McGinleyCooper}. The fact that after the quench the $\mathbb{Z}_2$ invariant (from Eq.~\eqref{Z2def}) for the BHZ model, which conserves $s_z$, remains well-defined and equal to the initial one is a consequence of the additional inversion symmetry.
\begin{figure}[h]
	\centerline{\includegraphics[width=125pt]{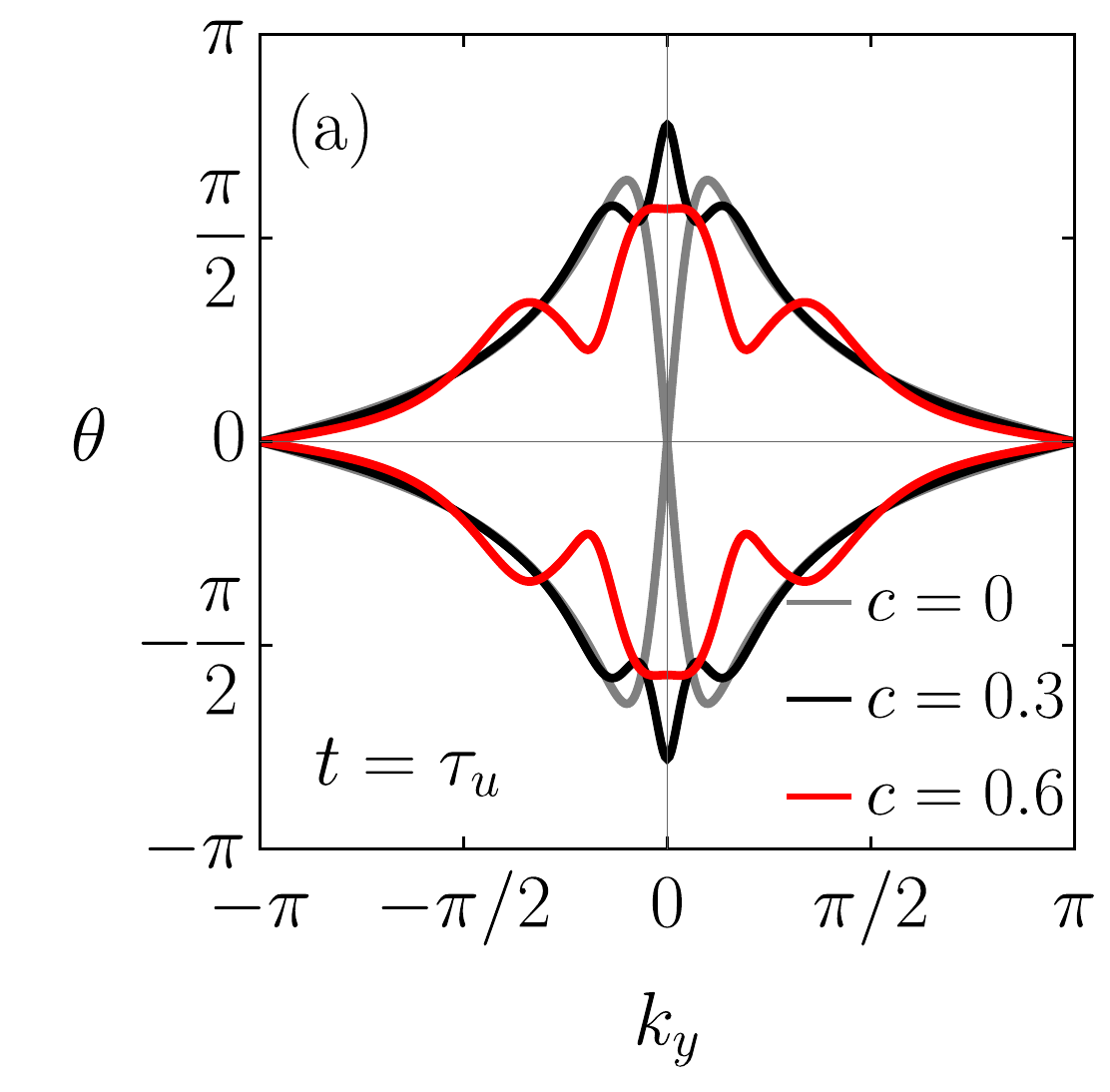}\includegraphics[width=125pt]{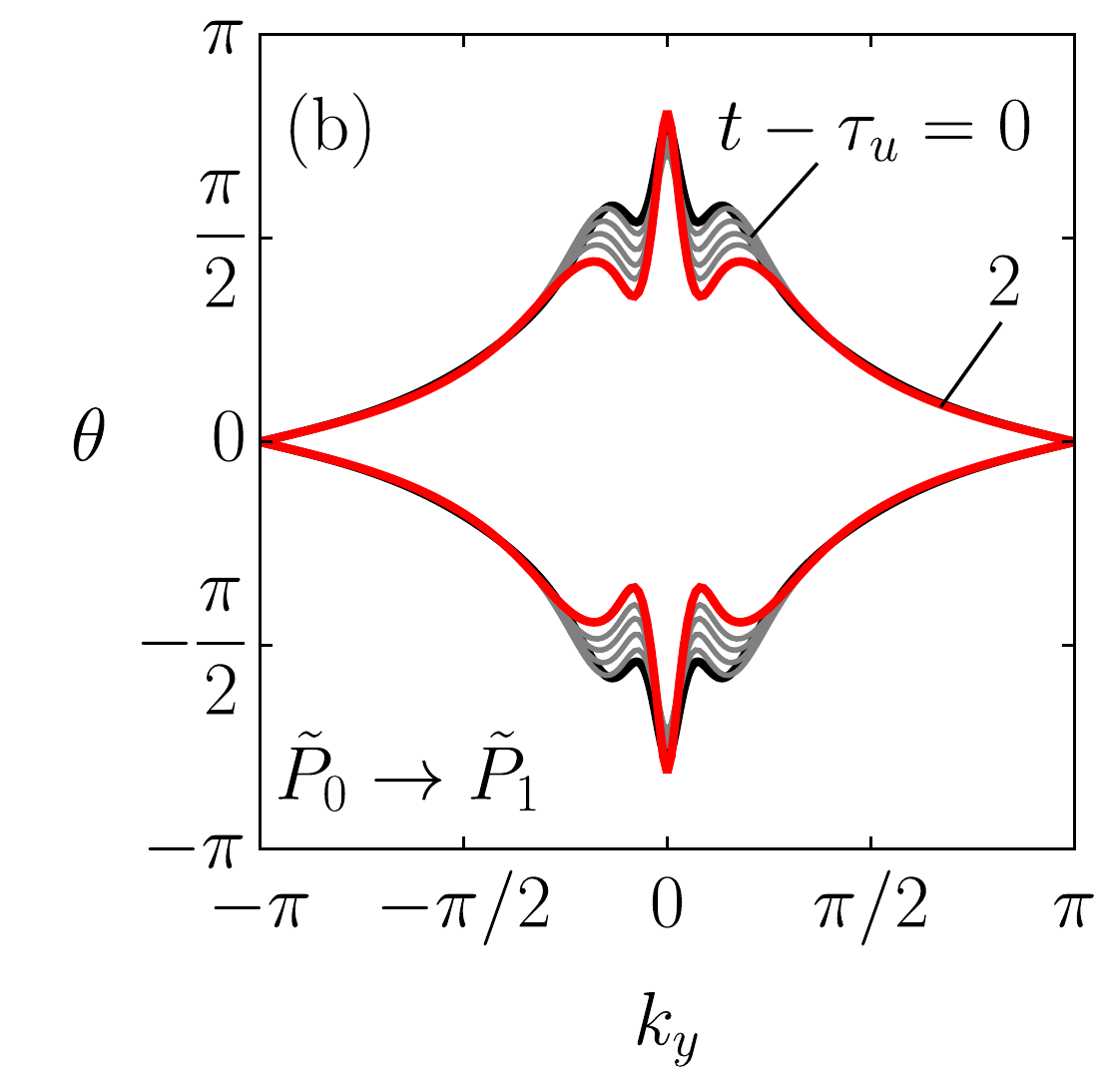}}
	\caption{(a) Wannier centre flow of systems with $c=0$ (grey), $c=0.3$ (black) and $c=0.6$ (red) at the end of the quench  from $u_0=-3$ to $u_1=-1$ with quench time $\tau_u=15$. (b) Wannier centre flow at $t=\tau_u$ (black), $t=\tau_u+2$ (red) and times in between (grey) after the $\tilde{P}_0\to\tilde{P}_1$ quench.}
	\label{fig:WanC}
\end{figure}
The Wannier centre flow of the time-evolved state for $c=0$ is shown in Fig.~\ref{fig:WanC}(a) (grey). The Wannier centre flow stays doubly degenerate at $k_y=0$ and $k_y=\pi$ for all times which is required for the calculation of the $\mathbb{Z}_2$ invariant according to Eq.~\eqref{Z2def}. The double degeneracy of Wannier centre flow is present due to inversion symmetry of the system (which is not broken by time evolution). Inversion symmetry constrains the Wannier centre flow at time-reversal symmetric momenta $K_y$ to $0$, $\pi$ or in the case of multi-band system to pairs with different sign $\theta_1(K_y)=-\theta_2(K_y)$ \cite{WilsonLoopBernevig}. As the two occupied bands of the BHZ model correspond to two independent Chern insulators, the Wannier centres can only take values $0$ or $\pi$ at $K_y$. Since the system evolves smoothly under Schr\"{o}dinger equation, the Wannier centre flow $\theta(K_y)$ cannot jump from $0$ to $\pi$, therefore it stays pinned to the initial value for all times.

Alternatively, for $c=0$ one could define the $\mathbb{Z}_2$ invariant also from the difference of the Chern numbers \cite{HasanKane10}. This quantity is conserved by the quench due to the conservation of the Chern numbers, and neither the inversion symmetry nor the limitation to two occupied bands is necessary in this case.
 
These considerations do not apply to systems with $c\neq0$ that do not conserve $s_z$ and hence to the general case with TRS only. Fig.~\ref{fig:WanC}~(a) shows Wannier centre flows of systems for several values of $c$ after the quench from a trivial to a topological regime. Contrary to the case with $c=0$, Wannier centre flows of the systems with $c\neq 0$ are not degenerate at $k_y=0$ which renders the $\mathbb{Z}_2$ invariant ill-defined.  The fact that the Wannier centre flow for the two bands takes opposite values at $k_y =0$ is a manifestation of the inversion symmetry that is still present in the state. After the quench, the Wannier centre flow becomes time-dependent. An example of time evolution of Wannier centre flow after the $\tilde{P}_0\to\tilde{P}_1$ quench is shown in Fig.~\ref{fig:WanC}~(b).

\section{Perturbative evaluation of the spin Hall conductivity}
\label{app:perturb}
\begin{figure}[h]
	\centerline{\includegraphics[width=125pt]{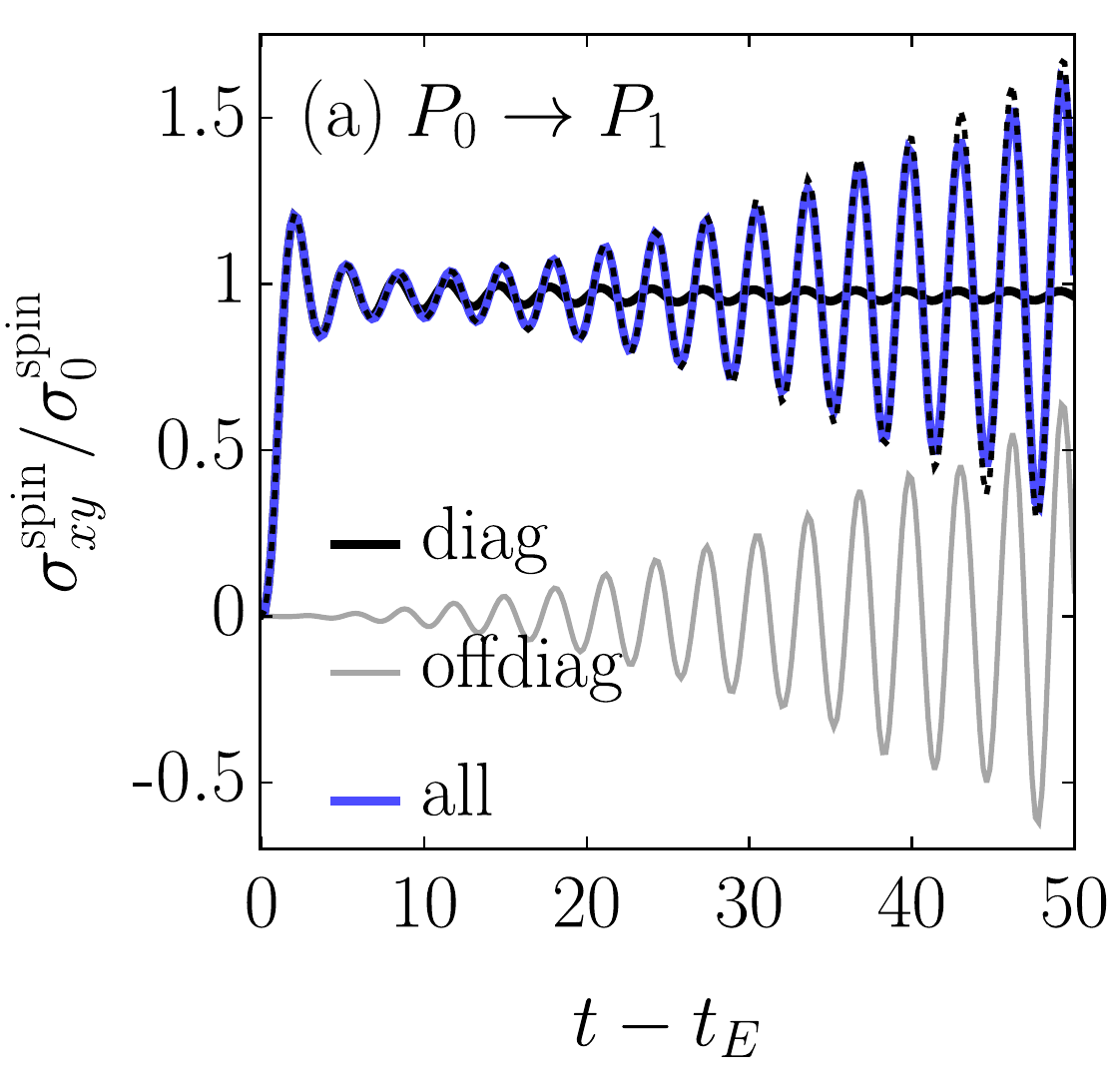} \includegraphics[width=125pt]{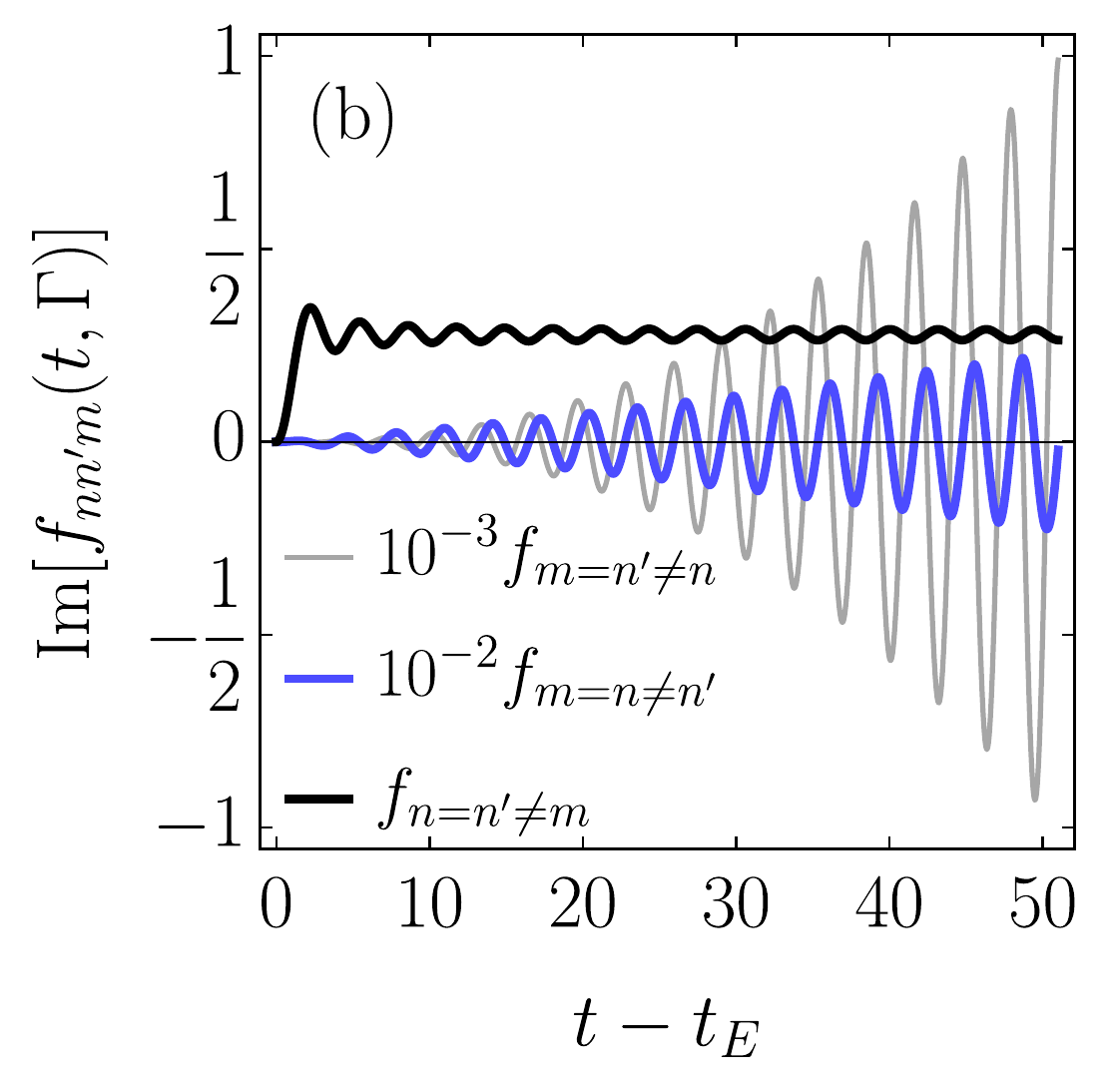}}
	\caption{ (a) Diagonal (black), off-diagonal (blue), all terms (red) in $\sigma_{xy}^\mathrm{spin}(t)$ according to Eq.~\eqref{eq:SpinHallConductivity} and $\sigma_{xy}^\mathrm{spin}(t)$ resulting from the full evaluation of Eq.~\eqref{eq:SpinHallConduct} (black, dotted) for the system after the $P_0\to P_1$ quench for $\tau_u =15$. (b) Imaginary part of the time-dependent element $f_{nn'm}(t,\Gamma)$ at the $\Gamma$ point, for the system at parameters $P_1$, is shown for different index combinations. In both figures the electric field is turned on as $E_x(t)=E_0[1-\exp (-(t-t_E)/\tau_E)]$.}
	\label{fig:DiagOffDiag}
\end{figure}

In the main text, we calculated the spin Hall conductivity from the expectation value of the spin current density \eqref{eq:SpinHallConduct}, which we evaluated for a Hamiltonian with explicitly included electric field. For small electric fields, one can evaluate the spin Hall conductivity also perturbatively. Let the electrons immediately  after the quench occupy the states $\{|\varphi_\alpha (\mathbf{k})\rangle=\sum_{n}c_{\alpha,n}(\mathbf{k})|u_n(\mathbf{k})\rangle,\,1\leq \alpha \leq N_F\}$ where $|u_n(\mathbf{k})\rangle$ are the eigenstates of the post-quench Hamiltonian $\hat{H}(\mathbf{k})$ without electric field. The response due to the electric field $E_x(t)$ can be evaluated using the time-dependent perturbation theory, as in Ref.~\onlinecite{Caio2016}. The resulting spin Hall conductivity reads
\begin{align}
\begin{aligned}
&\sigma_{xy}^\mathrm{spin}(t)=\frac{2e}{(2\pi)^2}\,\mathrm{Re} \sum_{\alpha}^{N_F}\sum_{n,n'm=1}^{2N_F}\intop\mathrm{d}\mathbf{k}\,c^{*}_{\alpha,n}(\mathbf{k})c_{\alpha,n'}(\mathbf{k})\times \\
&f_{nn'm}(t,\mathbf{k})\langle u_n(\mathbf{k})|\frac{1}{2}\hat{s}_z\partial_{k_y}\hat{H}|u_m(\mathbf{k})\rangle
\langle u_m(\mathbf{k})|\partial_{k_x}\hat{H}|u_{n'}(\mathbf{k})\rangle,\label{eq:SpinHallConductivity}
\end{aligned}
\end{align}
\begin{equation}
f_{nn'm}(t,\mathbf{k})=-i\, e^{i\triangle_{nm}(\mathbf{k})t}\intop_0^te^{i\triangle_{mn'}(\mathbf{k})t'}A_x(t')\,\mathrm{d}t'/E_x(t),
\end{equation}
where $\triangle_{nm}(\mathbf{k})=E_n(\mathbf{k})-E_m(\mathbf{k})$ and $A_x(t)=-\intop E_x(t)\mathrm{d}t$. In Eq.~\eqref{eq:SpinHallConductivity} the spin Hall conductivity is expressed with time-independent coefficients of the post-quench state $c_{\alpha,n}(\mathbf{k})$, likewise time-independent matrix elements of $\frac{1}{2}\hat{s}_z\partial_{k_y}\hat{H}$ and $\partial_{k_x}\hat{H}$ and a time-dependent function $f_{nn'm}(t,\mathbf{k})$ which is expressed in terms of energies of the states $E_n(\mathbf{k})$ and the dependence of the electric field on time. It turns out that for parameters used in our paper, Eq.~\eqref{eq:SpinHallConductivity} gives results that are essentially the same as that of the full evaluation specified in the main text. A comparison between the results of Eq.~\eqref{eq:SpinHallConductivity} (red line) and of the full evaluation (black, dotted line) is shown in Fig.~\ref{fig:DiagOffDiag} (a).

Expression \eqref{eq:SpinHallConductivity} is convenient for the interpretation of the results. The contributions to the spin Hall conductivity that include diagonal elements of the density matrix in the basis of Hamiltonian eigenstates $|c_{\alpha,n}(\mathbf{k})|^2$ (hereafter referred to as "diagonal terms"), and contributions  including off-diagonal elements of the density matrix $c_{\alpha,n}^{*}(\mathbf{k})c_{\alpha,n'}(\mathbf{k})$ (referred to as "off-diagonal terms") behave differently, see  Fig.~\ref{fig:DiagOffDiag} (a). One can see that the diagonal terms give rise to the finite average value of $\sigma_{xy}^\mathrm{spin}$. The frequency of the decaying small oscillations around this average value is given by the magnitude of the gap. The off-diagonal terms, on the other hand, exhibit oscillations around 0 with the magnitude that for long times increases in time quadratically. The observed time-dependence can be considered analytically by evaluating the time-dependent function $f_{nn'm}(t,\mathbf{k})$. 

The diagonal terms contain $\langle u_n(\mathbf{k})|\frac{1}{2}\hat{s}_z\partial_{k_y}\hat{H}|u_m(\mathbf{k})\rangle
\langle u_m(\mathbf{k})|\partial_{k_x}\hat{H}|u_{n}(\mathbf{k})\rangle$, which is non-vanishing only for index combinations $n\neq m$, where it is imaginary. Therefore, only the imaginary part of the function $f_{nnm}(t,\mathbf{k})$ contributes to the integral. The explicit evaluation for long times ($t\gg\tau_E$) yields  $\mathrm{Im}[f_{nnm}(t,\mathbf{k})]$
\begin{equation}
=\frac{1}{\triangle^2_{nm}(\mathbf{k})}-\frac{\cos[\triangle_{nm}(\mathbf{k})t]+\triangle_{nm}(\mathbf{k})\tau_E\sin[\triangle_{nm}(\mathbf{k})t]}{\triangle_{nm}(\mathbf{k})^2[1+\triangle_{nm}^2(\mathbf{k})\tau_E^2]}. 
\end{equation}
For long times, the function at every $\mathbf{k}$  oscillates around a finite mean value 
$1/\triangle_{nm}(\mathbf{k})^2$. The amplitude of oscillations vanishes with $\tau_E\to\infty$.

We now turn to the off-diagonal terms. The leading off-diagonal terms are given by  index combinations $m=n'\neq n$ and the corresponding time-dependent function for long times grows quadratically with $t$, $f_{nmm}(t,\mathbf{k})=i\,e^{-i\triangle_{nm}(\mathbf{k})t}[(t-\tau_E)^2+\tau_E^2]/2$. Other off-diagonal index combinations give rise to oscillations that grow linearly with time and are hence important only initially.
Fig.~\ref{fig:DiagOffDiag} (b) shows the imaginary part of the function $f_{nn'm}(t,\mathbf{k})$ for different index combinations.

Finally, we note that in the limit of an infinitely slow quench, excitations are present with probability $1$ only at $\mathbf{k}$ where the band gap closes. For such a system, the off-diagonal elements of the density matrix  $c_{\alpha,n}^{*}(\mathbf{k})c_{\alpha,n'}(\mathbf{k})$ are equal to zero for every $\mathbf{k}$, meaning that the growth of oscillations never occurs and the spin Hall response is equal to the ground-state one.

\bibliography{bibliography_list}{}

\end{document}